\documentclass[journal]{IEEEtran}
\usepackage{booktabs} 
\usepackage[utf8]{inputenc}
\usepackage{amsfonts,amssymb}
\usepackage{subfigure}
\usepackage{graphicx}
\usepackage{float}
\usepackage{multirow}
\usepackage{multicol}
\usepackage{blindtext}
\usepackage{color,colortbl}
\usepackage{amsmath}
\usepackage{cite}

\definecolor{Gray}{gray}{0.9}

\ifCLASSINFOpdf
\else
\fi

\newcommand{\wx}[1]{\textcolor{black}{#1}}

\begin{document}
%
\title{HDR-GAN: HDR Image Reconstruction from Multi-Exposed LDR Images with Large Motions}
%
%
%

\author{Yuzhen~Niu, \ \
        Jianbin~Wu, \ \
        Wenxi~Liu, \ \
        Wenzhong~Guo, \ \
        Rynson W.H. Lau
\thanks{Yuzhen Niu, Jianbin Wu, Wenzhong Guo, and Wenxi Liu are with the College of Mathematics and Computer Science, Fuzhou University, China. Wenxi Liu is the corresponding author.}
\thanks{Rynson W.H. Lau is with the Department of Computer Science, City University of Hong Kong.}
}

%
%

\markboth{Journal of \LaTeX\ Class Files,~Vol.~14, No.~8, August~2015}%
{Shell \MakeLowercase{\textit{et al.}}: Bare Demo of IEEEtran.cls for IEEE Journals}
%



\maketitle

\begin{abstract}
Synthesizing high dynamic range (HDR) images from multiple low-dynamic range (LDR) exposures in dynamic scenes is challenging. There are two major problems caused by the large motions of foreground objects. One is the severe misalignment among the LDR images. The other is the missing content due to the over-/under-saturated regions caused by the moving objects, which may not be easily compensated for by the multiple LDR exposures.
Thus, it requires the HDR generation model to be able to properly fuse the LDR images and restore the missing details without introducing artifacts. To address these two problems, we propose in this paper a novel GAN-based model, \textit{HDR-GAN}, for synthesizing HDR images from multi-exposed LDR images. To our best knowledge, this work is the first GAN-based approach for fusing multi-exposed LDR images for HDR reconstruction.
By incorporating adversarial learning, our method is able to produce faithful information in the regions with missing content.
In addition, we also propose a novel generator network, with a reference-based residual merging block for aligning large object motions in the feature domain, and a deep HDR supervision scheme for eliminating artifacts of the reconstructed HDR images. Experimental results demonstrate that our model achieves state-of-the-art reconstruction performance over the prior HDR methods on diverse scenes.
\end{abstract}


%
\IEEEpeerreviewmaketitle

\section{Introduction}

Although the human visual system has a much higher visual dynamic range, most off-the-shelf digital cameras typically produce photos with a limited range of illumination, which may not be desirable in many scenarios. Some specialized hardware devices~\cite{Nayar2000High,Tumblin2005Why} have been proposed to produce high dynamic range (HDR) images directly, but they are usually too expensive to be widely used.
In recent years, with prevailing mobile devices, there is a high demand for capturing HDR images of scenes using light-weighted and low-cost monocular mobile cameras, in order to produce photos that cover a broad range of illumination.

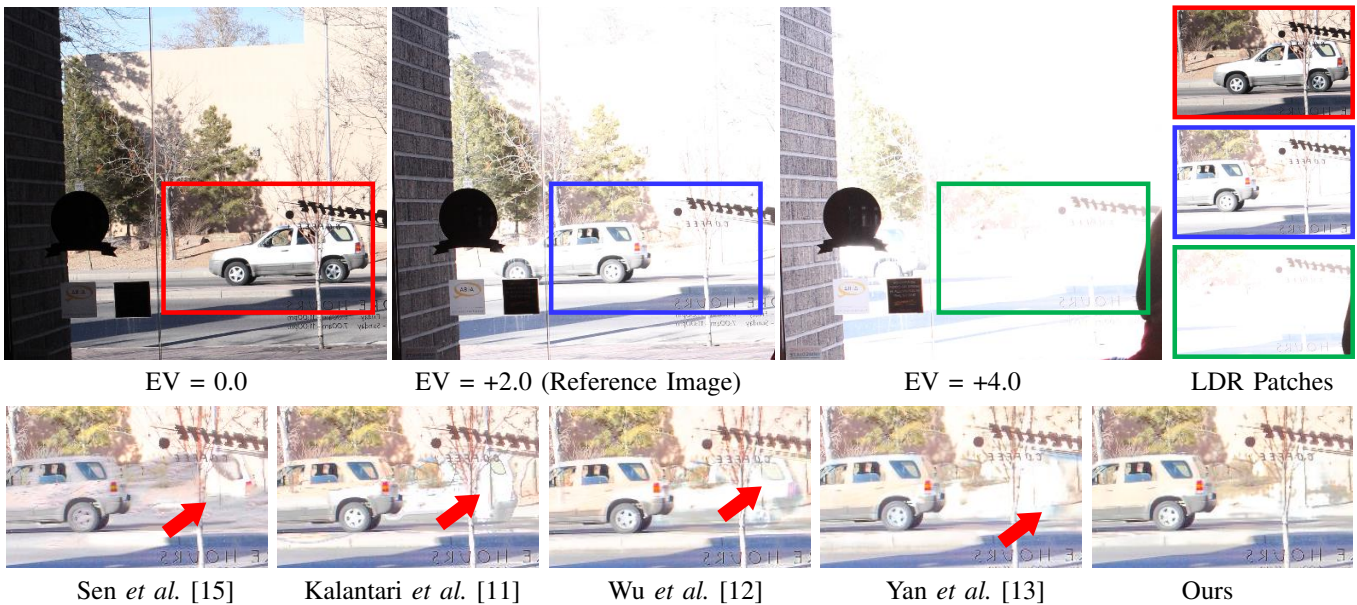
\begin{figure*}[t]
	\centering
	\includegraphics[width=\textwidth]{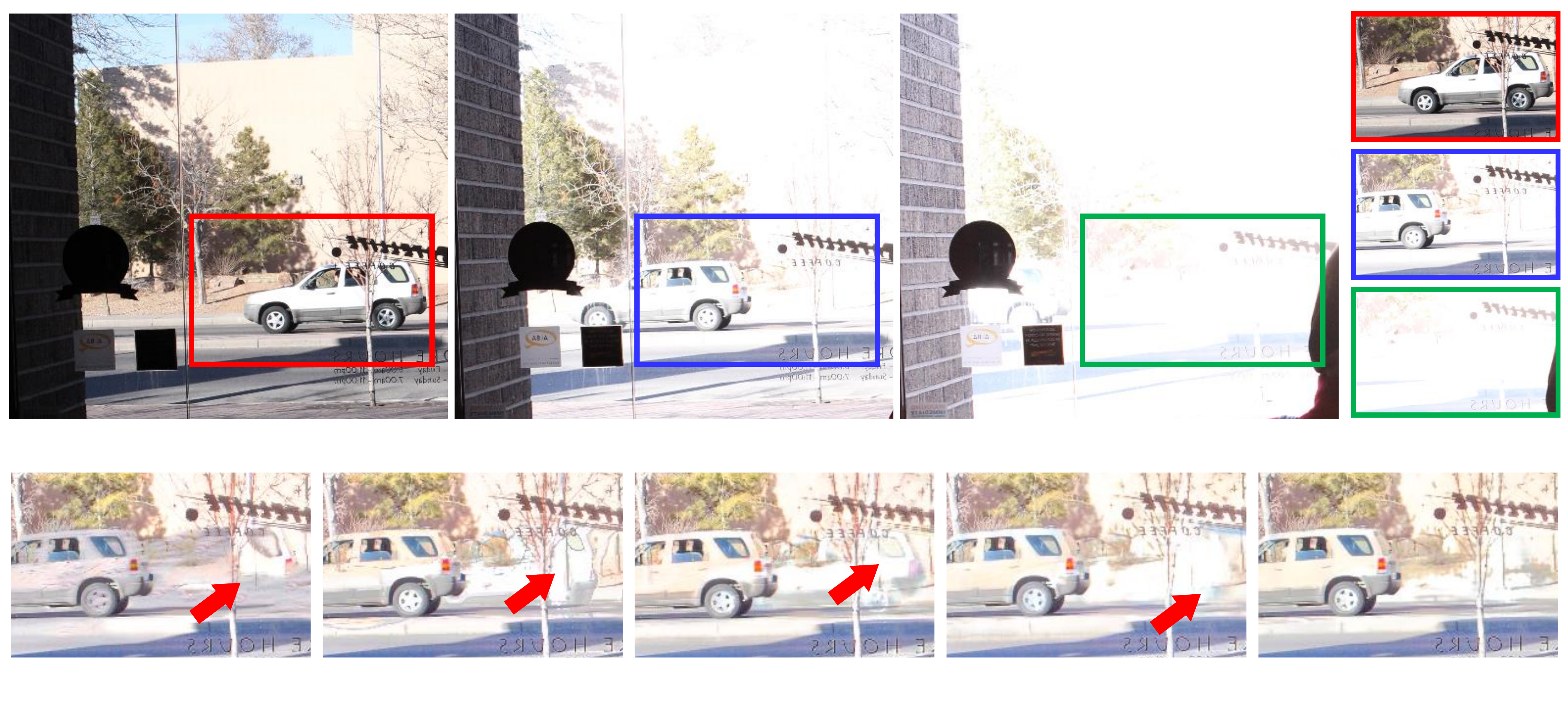}
	\begin{picture}(0,0)
	\put(-200,97){EV = 0.0}
	\put(-98,97){EV = +2.0 (Reference Image)}
	\put(87,97){EV = +4.0}
	\put(192,97){ LDR Patches }
	\put(-226,18){Sen \textit{et al.}~\cite{sen2012}}
	\put(-140,18){Kalantari \textit{et al.}~\cite{LearningHDR_Kalantari}}
	\put(-25,18){Wu \textit{et al.}~\cite{Wu_2018_ECCV_dhdr}}
	\put(80,18){Yan \textit{et al.}~\cite{gong2019attenhdr}}
	\put(192,18){Ours}
	\end{picture}
	\vspace{-0.8cm}
	\caption{
		Our GAN-based model is able to fuse multiple LDR images with large object motions into a ghost-free HDR image, without the need to explicitly align the images.
		In this example, a car is moving rapidly from right to left.
		In the over-exposure of the third shot (i.e., the green patch), the region where the car used to be in the first shot (i.e., the red patch) is dis-occluded. Compared with the patch-based method~\cite{sen2012} and recent deep learning based methods~\cite{LearningHDR_Kalantari,Wu_2018_ECCV_dhdr,gong2019attenhdr}, which all suffer from various degrees of ghosting artifacts (highlighted by the red arrows), our approach is able to restore the missing details and produce a high-quality HDR image.}
	\label{fig:teaser}
\end{figure*}

One popular approach to obtain an HDR image is to merge several low dynamic range (LDR) images captured from multiple exposures~\cite{Debevec1997,Mann95onbeing,Granados2010Optimal,Reinhard2005High}. Given a set of multi-exposed LDR images, one of them (typically with the medium exposure) is used as the reference image and the rest are used to compensate for the missing details due to over-/under-exposure of some local regions. When the LDR images are perfectly aligned pixel-wisely, these \wx{traditional methods that rely on hand-crafted features} can produce high quality HDR images. However, foreground and background misalignments are difficult to avoid in practice due to object motion, causing the output HDR images to have blur and ghosting artifacts \cite{Myszkowski2008High,ZimmerFreehand}.
In particular, there are two main problems caused by the large motions of foreground objects. One is the severe misalignment among the LDR images, and the other is the difficulty in compensating the missing content in the over-/under-exposed regions via multiple exposures due to occlusions of the moving objects in the images, as shown in Fig.~\ref{fig:teaser}.
%
%

\wx{To address these problems, some traditional methods attempt to align the LDR images in the preprocessing stage \cite{Grosch06fastand,Jacobs2008Automatic} before fusing them, e.g., using optical flow \cite{Jacobs2008Automatic,Myszkowski2008High}.}
\wx{With the popularity of deep learning, CNN-based HDR methods are proposed~\cite{LearningHDR_Kalantari,Wu_2018_ECCV_dhdr,gong2019attenhdr,yan2020deep}. These methods have strong ability in hallucinating details for the reconstructed HDR images. However, their network structures do not specifically handle the misalignment of the LDR images. For example, Kalantari \textit{et al.}~\cite{LearningHDR_Kalantari} directly utilize a CNN for pixel-wise merging. Wu \textit{et al.}~\cite{Wu_2018_ECCV_dhdr} apply the U-Net or ResNet architecture with sparse skip connections for LDR fusion. Yan \textit{et al.}~\cite{gong2019attenhdr} introduces dense connections into the model, without considering multi-scale context information of the LDR images. These methods cannot guarantee that the fused features of the LDR images are aligned well.
Thus, they still require to incorporate homography transformation or optical flow to holistically align the LDR images beforehand, to cover the shortage of their networks.
However, when the LDR images contain large motions or significant misalignment, they may still suffer from having artifacts due to the unreliability of optical flow estimation, especially for images captured with different exposure levels, as demonstrated in Fig.~\ref{fig:teaser}. }

In this work, we address the two aforementioned problems by proposing a novel GAN-based model, \textit{HDR-GAN}, which produces high-quality HDR images from multi-exposed LDR images without the need to explicitly align the LDR images.
%
%
To our best knowledge, our approach is the first attempt to adapt a GAN-based model in this task for fusing multi-exposed LDR images to reconstruct an HDR image, which can help recover regions with incomplete image content due to under-/over-exposures.
To address the limitations of prior CNN-based methods in fusing the misaligned LDR features, in our GAN-based framework, we propose a reference-based residual merging module for aligning large object motions in the feature domain and a deep HDR supervision scheme for eliminating artifacts in the reconstructed HDR image.
\wx{In particular, the proposed reference-based residual merging module contains
reference-based residual structures to transform the LDR features of multiple exposures to compensate for the reference image and thus implicitly align the features of low and high exposures to that of the median exposure (i.e., the reference image).}
\wx{In addition, inspired by the deep supervision of prior models~\cite{szegedy2015going,zhou2018unet++,sun2019deep}, we introduce a deep HDR supervision scheme into our model. We first upsample the merged features of each scale, and sequentially pass them into the decoding block of its upper scale until reaching the top scale to generate several high-resolution HDR images for supervision. Supervising the resulting HDR images can then encourage the feature fusion of different scales, while eliminating the artifacts caused by the misalignment of the LDR image features.}

To further encourage high-quality HDR reconstruction under challenging scenarios with severe missing details, we leverage the adversarial loss from \cite{Park_2019_CVPR_sphere_gan} for discriminating the generated samples and the ground-truth by projecting them to a hypersphere space, which helps boost the image generation capability of our model. As demonstrated in Fig.~\ref{fig:teaser}, our model is able to hallucinate more details for the unobserved regions in the reference LDR image. To evaluate the performance of our model, we compare our approach with other HDR methods on a public benchmark and show that our model can achieve the state-of-the-art performance.

To sum up, the main contributions of our work include:
\begin{itemize}
    \item We propose the first GAN-based method, \textit{HDR-GAN}, for HDR reconstruction from multi-exposed LDR images. By incorporating adversial learning, our method is able to produce faithful information for the HDR images when LDR images contain large object motions.
    \item We propose a novel generator network, with a reference-based residual merging block to implicitly align large object/camera motions in the feature domain, and a deep HDR supervision scheme for eliminating artifacts.
    \item Our experimental results on the public benchmark demonstrate that the proposed model outperforms the state-of-the-art HDR models.
\end{itemize}



\section{related work}

In this section, we briefly summarize related work on HDR reconstruction for static and dynamic scenes.

\subsection{HDR Reconstruction for Static Scenes}

There are HDR reconstruction methods that assume the multi-exposed input LDR images are aligned. These methods are essentially handling a static scene, in which all objects do not change during the capturing period. They typically use hand-crafted features or deep features to merge the LDR images pixel-wisely \cite{barakat2008minimal, hasinoff2010noise, seshadrinathan2012noise, pourreza2015exposure, prabhakar2017deepfuse,beek2019improved}. 

In recent years, with the progress of deep learning techniques, researchers attempt to address a more challenging problem of reconstructing an HDR image from a single LDR image~\cite{EKDMU17,lee2018deep,Khan2019fhdr,Liu-Single-CVPR-2020}.
For example, Eilertsen \textit{et al.}~\cite{EKDMU17} utilize the U-Net architecture to learn the mapping from LDR images to HDR images.
Lee \textit{et al.}~\cite{lee2018deep} propose to generate LDR images with multiple exposures from a single LDR image using a GAN, and then merge the pseudo multi-exposed images pixel-wisely to produce an HDR image. Liu \textit{et al.}~\cite{Liu-Single-CVPR-2020} propose an HDR-to-LDR image formation pipeline that includes dynamic range clipping, non-linear mapping from a camera response function, and quantization. They then propose to learn three specialized CNNs to reverse these steps in order to reconstruct HDR images. Since it often exists missing details from a single image caused by quantization and saturation of the camera sensor, the performance of these single image based HDR reconstruction methods generally do not perform well, compared with methods based on multiple exposures.

\subsection{HDR Reconstruction for Dynamic Scenes}

While LDR images with multiple exposures contain rich information of the scene, it is difficult to avoid object motions during the image capture period. A number of methods have been proposed to address this problem~\cite{Grosch06fastand,Jacobs2008Automatic,Myszkowski2008High,sen2012,LearningHDR_Kalantari,Wu_2018_ECCV_dhdr,gong2019attenhdr,yan2019wacv,yan2020deep,chen2020deep}. Traditional HDR reconstruction methods for dynamic scenes estimate homography transformations to handle camera motions or to use optical flows as prior to align LDR images in the preprocessing stage~\cite{LearningHDR_Kalantari,Jacobs2008Automatic,Myszkowski2008High}.

Recent CNN-based methods formulate HDR reconstruction as an image translation problem from the LDR images to the HDR image~\cite{Wu_2018_ECCV_dhdr,gong2019attenhdr,yan2019wacv,yan2020deep}.
Wu \textit{et al.}~\cite{Wu_2018_ECCV_dhdr} first estimate the homography transformation to align the background of the LDR images, and then propose an U-Net or ResNet structure to directly learn the mapping from LDR images to a high-quality HDR image.
Yan \textit{et al.}~\cite{gong2019attenhdr} introduce spatial attention during the process of fusing LDR image features, while Yan \textit{et al.}~\cite{yan2020deep} leverage non-local blocks to process the global context of the unaligned image features.
However, there are two challenging issues caused by the large motions of foreground objects. One is the severe misalignment of the LDR images, while the other is that the missing contents in the over-/under-exposed regions may not be easily compensated using multiple exposures, caused by the dis-occlusion of the moving objects in the scene. These existing methods either cannot guarantee to align the LDR images or fail to produce adequately faithful information for the missing image contents. Recently, there is a concurrent work~\cite{yan2020deep} that explicitly addresses the feature alignment problem using a non-local network. However, due to the limited generation capability of the network structure, it is unable to hallucinate the missing contents caused by large object motions well.

In this work, we propose a GAN-based model to address the limitations of existing methods in handling large object motions in the scene.

\section{Our Proposed Method}
\label{sec:method}

\begin{figure*}
    \centering
    \includegraphics[width=0.9\textwidth]{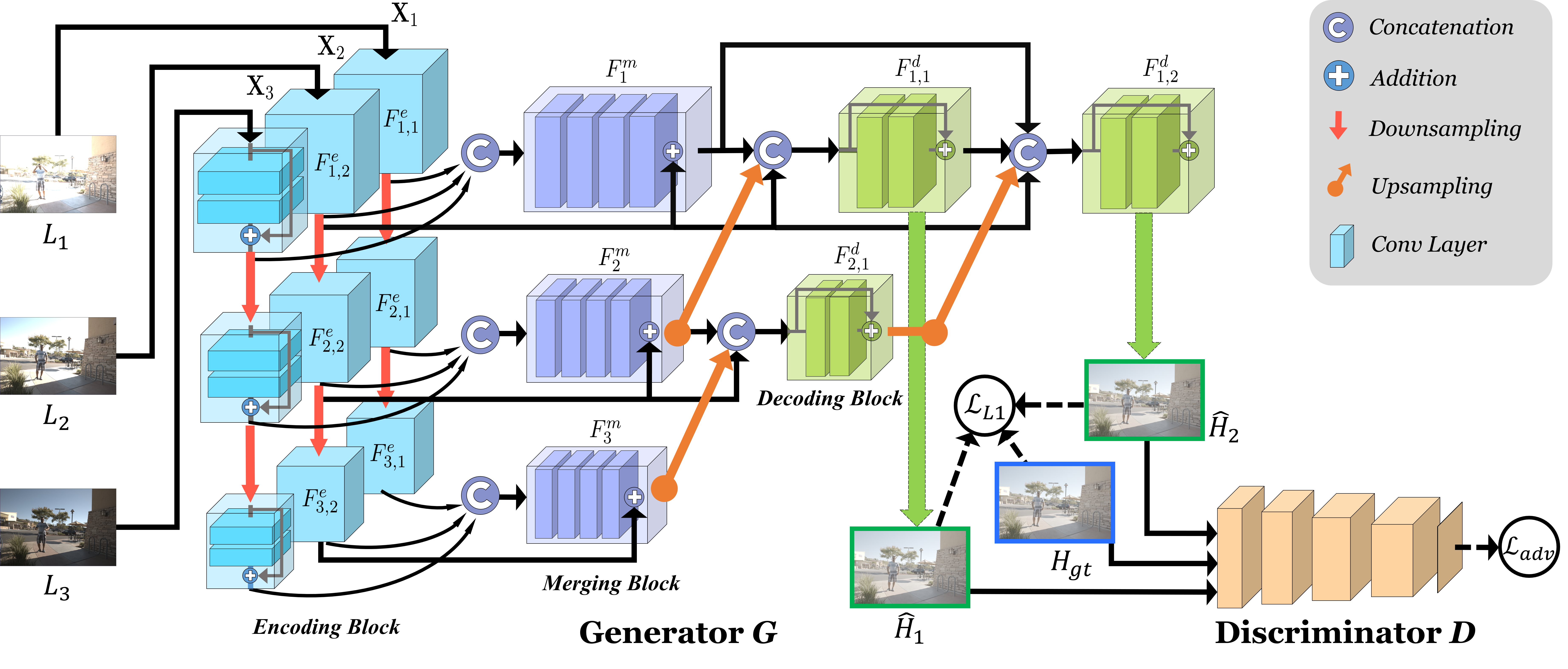}
    \vspace{-3mm}
    \caption{\textbf{Illustration of the proposed framework.} Our framework is composed of two main networks: the generator $G$ and the discriminator $D$. As illustrated, the three input LDR images $L_i$ ($i \in \{1,2,3\}$) are fed into the generator network $G$ to produce the output HDR image. $G$ consists of the encoding blocks (i.e., the blue cubes) that extract features from the LDR images, the merging blocks (i.e., the light purple cubes) that align the LDR features, and the decoding blocks (i.e., the green cubes) that restore the aggregated features to larger scales.
    It produces two HDR images $\hat{H}_1$ and $\hat{H}_2$ for supervision with L1 loss, while $\hat{H}_2$ represents the output HDR image.
    On the other hand, the discriminator network $D$ is employed to discriminate between the generated images and the ground-truth via an adversarial loss.}
    \label{fig:framewrok}
\end{figure*}

Fig. \ref{fig:framewrok} demonstrates the proposed HDR-GAN framework. Given three LDR images of different exposures (i.e., $L_1$, $L_2$ and $L_3$, sorted by their exposure biases) as input and $L_2$ as the reference image, our goal is to construct an HDR image. As in previous works \cite{LearningHDR_Kalantari,Wu_2018_ECCV_dhdr,gong2019attenhdr}, the input LDR images $L_i$ ($i=\{1,2,3\}$) are first converted to the HDR domain (i.e., the value of each pixel is mapped to the range of $[0,1]$), and then fused to produce an HDR image via the proposed deep model. Finally, \wx{following \cite{LearningHDR_Kalantari},} the loss is measured by passing the generated HDR image through a differentiable tonemapper (i.e., $\mu$-law), \wx{which is commonly used for range compression in audio processing:}
\begin{align}
	\mathcal{T}(H)=\frac{\log (1+\mu H)}{\log(1+\mu)},
	\label{eq:mu_law}
\end{align}
where $H$ represents an HDR image and $\mu$ controls the extent of compression.

The major challenge of fusing LDR images with different exposures to an HDR image is the presence of large motions in dynamic scenes, leading to the missing content and the misalignment among the LDR images. To encourage our model to hallucinate the missing details and thus to generate a high-quality HDR image, we propose to employ the GAN framework. In the past few years, GANs have been extensively studied and successfully applied to many image generation tasks.
Similar to the standard GAN framework~\cite{goodfellow2014generative,isola2017image}, our model consists of a generator $G$ and a discriminator $D$ competing against each other in a two-player min-max game. In our problem, the generator tends to merge the LDR images and produce a high-quality HDR image. The discriminator, on the other hand, aims to classify the generated image and the ground-truth image as real or fake.

In the following subsections, we introduce the technical details of our proposed generator and discriminator.

\subsection{The Generator}

The purpose of the generator is to
produce a high-quality HDR image $\hat{H}$, given the input LDR images $L_i$ ($i=\{1,2,3\}$). As shown in Fig. \ref{fig:framewrok}, we follow existing practice \cite{LearningHDR_Kalantari}, and feed the multi-exposed LDR images concatenated with their corresponding HDR versions as input $\textbf{X}_i$ ($i=\{1,2,3\}$) to the generator:
\begin{equation}
    \textbf{X}_i=\{L_i \oplus F_{gamma}(L_i)\}=\{L_i\oplus L_i^\gamma / t_i\},
    \label{eq:gamma-correct}
\end{equation}
where $F_{gamma}(L_i)$ denotes the HDR version of $L_i$, $\gamma$ denotes the Gamma correction parameter (with $\gamma=2.2$) and $t_i$ denotes the exposure time of $L_i$. $\oplus$ denotes the concatenation operator.
Hence, our generator network $G$ can be briefly formulated as $\hat{H} = G(\textbf{X}_1, \textbf{X}_2, \textbf{X}_3)$.

Due to the large motions of dynamic objects in the scene or the camera, the misalignment of the LDR images tend to produce misalignment of the extracted features, causing the model to generate undesired artifacts in the output HDR images.
To address this concern, as shown in Fig.~\ref{fig:framewrok}, we propose a novel generator network that is composed of specifically-designed modules, including multi-scale LDR encoding blocks and reference-aligned feature merging blocks, together with deep HDR supervision to enable implicitly alignment of the LDR images in the feature domain to reduce the artifacts of the resulting HDR images.

\textbf{Multi-scale LDR encoder.} For each input $\textbf{X}_i$, we apply an individual encoder to extract its features in multiple scales, such that patch-level features of the LDR images can be learned for later fusion. \wx{Although prior works~\cite{Wu_2018_ECCV_dhdr,gong2019attenhdr} also introduce individual encoders, they do not fully exploit the multi-scale context information obtained from the encoders. Both of them adopt shallow encoders with two convolutional layers only, without considering multi-scale. In contrast, our encoders have three scales to process each input $\textbf{X}_i$, and each scale is processed by one encoding block, i.e., a residual structure with two convolutional layers. The ablation study in Section~\ref{sec:ablation} demonstrates that our multi-scale LDR encoder performs the best.}

Specifically, each LDR encoder of our network contains three consecutive downsampling encoding blocks (represented by the blue cubes from the top to bottom of Fig.~\ref{fig:framewrok}), which help extract visual features in $1^{st}$, $2^{nd}$, and $3^{rd}$ scales from one input $\textbf{X}_i$. Formally, we denote the encoding blocks of different scales as $F_{j,i}^e$, $j,i=\{1,2,3\}$,  where $i$ refers to the index of the input $\textbf{X}$ and $j$ refers to the index of the network scale.
Each encoding block is a two-layer residual structure without the pooling layers, followed by a convolutional layer with a $3\times 3$ kernel size and stride $2$ for downsampling the feature maps (shown as red arrows in Fig.~\ref{fig:framewrok}). Hence, the encoded features at the $j$-th scale for $\textbf{X}_i$, denoted as $\textbf{E}_{j,i}$, is computed by sequentially passing $\textbf{X}_i$ through the encoding blocks from the top scale (i.e., $j=1$) to the bottom scale (i.e., $j=3$) as:
\begin{align}
    \textbf{E}_{1,i} &= F_{1,i}^e(\textbf{X}_i), \nonumber \\
    \textbf{E}_{2,i} &= F_{2,i}^e(F^{down}(F_{1,i}^e(\textbf{X}_i))), \nonumber \\
    \textbf{E}_{3,i} &= F_{3,i}^e(F^{down}(F_{2,i}^e(F^{down}(F_{1,i}^e(\textbf{X}_i))))),
\end{align}
where $F^{down}(\cdot)$ is implemented as a convolutional layer with stride $2$.

\textbf{Reference-aligned feature fusion.} After extracting the features from the input LDR images, we fuse the features of different scales using the proposed merging blocks as shown in Fig.~\ref{fig:framewrok}, which also help align the image content in the feature domain with a residual structure.
In particular, the encoded features from three encoding blocks of the same scale are concatenated and transformed by a merging block, i.e., $F_{j}^m$ ($j=\{1,2,3\}$), as:
\begin{align}
    \textbf{M}_j = F_j^m(\{\textbf{E}_{j,1}\oplus\textbf{E}_{j,2}\oplus \textbf{E}_{j,3}\}, \textbf{E}_{j,2}),
    \label{eq:merging_block}
\end{align}
where $\textbf{M}_j$ is the merged features of the $j^{th}$ scale. $\oplus$ denotes the concatenation operator. Hence, $\{\textbf{E}_{j,1}\oplus\textbf{E}_{j,2}\oplus \textbf{E}_{j,3}\}$ represents the concatenation of the encoded features from multiple exposures in the $j^{th}$ scale. In each merging block, we introduce a reference-based residual structure.
According to Eq.~\ref{eq:merging_block}, there is an additional input $\textbf{E}_{j,2}$ (i.e., the encoded features from the reference image) to the merging block $F_j^m$. $\textbf{E}_{j,2}$ is directly fed to the end of the merging block to form a residual structure. The purpose of this design is to enforce the alignment of the encoded features by utilizing the features from the three LDR images to compensate the features from the reference image (i.e., $\textbf{E}_{j,2}$). In essence, the features from the low and high exposures will be adapted to those of the median exposure (i.e., the reference image) to mitigate the misalignment problem.
Represented by light purple cubes in Fig. \ref{fig:framewrok}, each merging block is composed of four dilated convolutional layers with $3\times 3$ kernel size and a dilated factor of $2$ to transform the concatenated features and allow $\textbf{E}_{j,2}$ to skip-connect to the end of the block.

\textbf{Deep HDR supervision.} \wx{To further eliminate the artifacts in order to obtain high-fidelity results, we propose a deep HDR supervision scheme. Compared with prior deep supervision structures~\cite{szegedy2015going,zhou2018unet++,sun2019deep,ronneberger2015u}, we upsample the intermediate merged features and concatenate them with those in the upper scales to produce the HDR images at the original spatial dimension using a series of decoding blocks. With the aggregated cross-scale features and the skip-connected encoded features from the reference image involved in the decoders, as demonstrated in the ablation study in Section~\ref{sec:ablation}, the deep supervision can further enforce the alignment of the merged features and thus our network is able to produce better quality HDR images.}

Specifically, at the $j^{th}$ scale, we concatenate the features from the reference image (i.e., $\textbf{E}_{j,2}$), the merged features $\textbf{M}_j$ computed by Eq.~\ref{eq:merging_block}, and the upsampled features of the lower scale $F^{up}(\textbf{M}_{j+1})$ into feature maps with rich information. Note that $\textbf{E}_{j,2}$, $\textbf{M}_j$ and $F^{up}(\textbf{M}_{j+1})$ have all been aligned with the features from the reference image.

Recall that in the traditional encoder-decoder architecture (e.g., \cite{ronneberger2015u}), the encoded features of the deepest scale will be fed to the decoding blocks and upsampled sequentially.
To enhance the image generation quality, instead of supervising only the last decoding block  as the standard decoders do, we also want to control the fused intermediate features of the network. To do this, besides the standard upsampling path (i.e., $F_3^m \rightarrow F^d_{2,1} \rightarrow F^d_{1,2}$ in Fig.~\ref{fig:framewrok}), we introduce an auxiliary upsampling path (i.e., $F^m_2 \rightarrow F^d_{1,1}$ in Fig.~\ref{fig:framewrok}). While the standard upsampling path outputs an image $\hat{H}_2$ at the same dimension as the inputs, the auxiliary upsampling path aggregates the intermediate features and produces $\hat{H}_1$ also at the same size as the inputs.

Note that since the auxiliary upsampling path consists of fewer layers, it leads to faster convergence during training and its output is concatenated to the last decoding block, $F^d_{1,2}$.
This upsampling path not only enhances the quality of the output images, but also benefits network learning, i.e., $\hat{H}_2$ can be complimentary to the incomplete details of $\hat{H}_1$. In addition, the merged features at the top scale, $F^m_1$, is also consecutively passed through $F^d_{1,1}$ and $F^d_{1,2}$ for a thorough spatial transformation to further refine the features.
Hence, the computation of each block of the decoder can be expressed as:
\begin{align}
    \textbf{C}_{1,1} &= F_{1,1}^d(\textbf{M}_{1}\oplus \textbf{E}_{1,2}\oplus F^{up}(\textbf{M}_{2})), \nonumber \\
    \textbf{C}_{2,1} &= F_{2,1}^d(\textbf{M}_{2}\oplus \textbf{E}_{2,2}\oplus F^{up}(\textbf{M}_{3})), \nonumber \\
    \textbf{C}_{1,2} &= F_{1,2}^d(\textbf{M}_{1}\oplus \textbf{E}_{1,2}\oplus\textbf{C}_{1,1}\oplus  F^{up}(\textbf{C}_{2,1})),
\end{align}
where $\textbf{C}_{j,k}$ refers to the output of the decoding block $F_{j,k}^d$ ($k=\{1,\cdots,3-j\}$) and $F^{up}$ denotes the upsampling layer. For each scale, we concatenate the features computed from the preceding layers with the upsampled features from the smaller scale to strengthen the feature representation.
Besides, we also introduce the features from $\textbf{X}_2$ (i.e.,  $\textbf{E}_{j,2}$) that contains the reference image with median exposure to join the decoding blocks via skip connections.
Furthermore, as shown in the green cubes of Fig.~\ref{fig:framewrok}, each decoding block has an identical residual structure as the encoding blocks, which is followed by a nearest-neighbor interpolation layer.

For training, our model generates two HDR images $\hat{H}_1$ and $\hat{H}_2$ at the two output branches $\textbf{C}_{1,1}$ and $\textbf{C}_{1,2}$. Both are supervised by the ground-truth HDR image $H_{gt}$ via an $L_1$ loss as:
\begin{equation}
    \mathcal{L}_{L1} = \min_G (||\mathcal{T}(\hat{H}_1)-\mathcal{T}(H_{gt})||_1 + ||\mathcal{T}(\hat{H}_2)-\mathcal{T}(H_{gt})||_1),
    \label{eq:loss-mu}
\end{equation}
where $\mathcal{T}(\cdot)$ is the differentiable tonemapper defined in Eq.~\ref{eq:mu_law}.
Thus, our deep HDR supervision scheme is able to eliminate the artifacts and improve the quality of the reconstructed HDR images.

\subsection{The Discriminator}

We adopt PatchGAN \cite{isola2017image}, which contains five convolutional layers, as the discriminator $D$.
We also adopt the adversarial loss of \cite{Park_2019_CVPR_sphere_gan} for stabilizing the training process and avoiding modal collapse. Specifically, the output of the discriminator can be reshaped as an $n$-dimensional vector, i.e., $\textbf{q} = D(\cdot) \in \mathbb{R}^n$, and then projected to a point $\textbf{p}$ on a hypersphere $\mathbb{S}^n$ by the inverse of the stereographic projection, as:
\begin{align}
    \textbf{p} = \left(\frac{2 \textbf{q} }{||\textbf{q}||^2+1}, \frac{||\textbf{q}||^2-1}{||\textbf{q}||^2+1}\right).
    \label{eq:inv-ste-project}
\end{align}
Thus, instead of measuring the Euclidean distance between two points $\textbf{q}$ and $\textbf{q}^\prime$, we can measure the distance of any two projected points on the hypersphere by $d_s(\textbf{p},\textbf{p}^\prime)$ as:
\begin{align}
    d_s(\textbf{p},\textbf{p}^\prime) = \arccos{\frac{||\textbf{q}||^2||\textbf{q}^\prime||^2- ||\textbf{q}||^2- ||\textbf{q}^\prime||^2+4\textbf{q}\textbf{q}^\prime+1}{(||\textbf{q}||^2+1)(||\textbf{q}^\prime||^2+1)}}.
\end{align}
In this GAN-based formulation, we can optimize the discriminator based on the distances of the mapping features from the generated images and those from the ground-truth images, with respect to the reference point $\textbf{N} = [0, ..., 0, 1]^T \in \mathbb{R}^n$, i.e., the north pole of the hypersphere.
The formulation can be expressed as:
\begin{align}
    \mathcal{L}_{adv} = &\min_G \max_D \sum_{r} \mathbb{E}_{\textbf{z}}\left[d^r_s\left(\textbf{N}, D(\textbf{z})\right)\right]  \nonumber \\
    &- \sum_r \mathbb{E}_{\textbf{x}_1,\textbf{x}_2,\textbf{x}_3} \left[d^r_s\left(\textbf{N}, D\left( G(\textbf{x}_1,\textbf{x}_2,\textbf{x}_3)\right)\right)\right],
    \label{eq:adv_loss}
\end{align}
where $G(\cdot)$ and $D(\cdot)$ refer to the generator and discriminator, respectively.
$d^r_s(\cdot,\cdot)$ measures the $r^{th}$ moment distance (we set $r=\{1,2,3\}$) on the hypersphere between $D(\cdot)$ and the north pole of the hypersphere (i.e., $\textbf{N}$). In addition, $\textbf{z}$ is sampled from the distribution of the ground-truth images, while $\textbf{x}_i$ is sampled from the LDR images.
To sum up, we train our \textit{HDR-GAN} with the following hybrid loss:
\begin{equation}
    \mathcal{L} = \mathcal{L}_{L1} + \lambda \cdot \mathcal{L}_{adv},
\end{equation}
where ${\lambda}$ is a predefined constant that balances the loss terms, and we set it to 1 in our implementation.

\section{Experimental Results}
\label{sec:exp}

\subsection{Datasets and Metrics}
To evaluate the proposed method, we use the HDR dataset \cite{LearningHDR_Kalantari} for training and evaluation. It contains 74 image sets for training and 15 image sets for testing. For each training image set, three different LDR images are captured with exposure biases of $\{-2, 0, +2\}$ or $\{-3, 0, +3\}$. We also test our model on the datasets without ground-truth, including Sen's and Tursun's datasets~\cite{sen2012,tursun2016}.

We use PSNR and SSIM as the metrics for evaluation. Specifically, we compute PSNR$_\mu$ and SSIM$_\mu$ between the fused HDR image (i.e., $\hat{H}_2$) and the ground truth image, both in the tonemapping domain by $\mu$-law (Eq.~\ref{eq:mu_law}) with $\mu = 5000$, which is consistent with the setting of the prior works \cite{LearningHDR_Kalantari,Wu_2018_ECCV_dhdr,gong2019attenhdr}. We also compute PSNR$_L$ and SSIM$_L$ for comparison in the linear domain (i.e., HDR domain). Further, we compute the HDR-VDP-2 \cite{hdr_vdp}, which assesses the visibility and quality of the HDR images in different luminance conditions. Following \cite{Wu_2018_ECCV_dhdr}, for the two parameters used to compute the HDR-VDP-2 scores, we set the diagonal display size to 24 inches, and the viewing distance to 0.5 meter.

\subsection{Implementation and Details}

We have implemented our model using Tensorflow and evaluated it on an NVIDIA Tesla V100 GPU with 16G memory. We adopt spectral normalization \cite{miyato2018spectral} instead of batch normalization in our model.
All training images are at $1500 \times 1000$ resolution. Specifically, we sample patches of size $512 \times 512$ from the training set for training. To augment the training set, we randomly crop the training images and apply random rotation/flipping on the patches.
The learning rate of our network is $10^{-4}$ initially, decays to $10^{-5}$ after 114,300 iterations, and then drops to $10^{-6}$ after 706,400 iterations. We train our model with a batch size of $2$ for 1,215,000 iterations in total using the Adam solver, and the entire training process takes around a week. During inference, our model takes on average 0.29s and 8,833 MB memory to process a set of LDR images at $1440\times 960$ resolution.

\subsection{Comparison with the State-of-the-Arts}

\begin{table}
	\centering
	\footnotesize
	\setlength{\tabcolsep}{4pt}
	\caption{Comparison with the state-of-the-art methods. The best performance values are indicated in bolded. %
	\(^\dagger\) indicates that the results are from \cite{Wu_2018_ECCV_dhdr} and \cite{yan2020deep}. (All these results are the higher the better.)}
	\vspace{-0.3cm}
	\begin{tabular}{c||ccccc}
\toprule
Method & PSNR$_\mu$ & PSNR$_L$ & SSIM$_\mu$ & SSIM$_L$ & HDR-VDP-2 \\
\midrule
Sen \textit{et al.}~\cite{sen2012}\(^\dagger\) & 40.800 & 38.110 & 0.9808 & 0.9721 & 59.38 \\
Hu \textit{et al.} \cite{hu2013hdr}\(^\dagger\) & 35.790  & 30.760 & 0.9717 & 0.9503 & 57.05 \\
Kalantari \textit{et al.}~\cite{LearningHDR_Kalantari} & 42.670 & 41.232 & 0.9888 & 0.9846 & 65.05 \\
DeepHDR \cite{Wu_2018_ECCV_dhdr}\(^\dagger\) & 41.650  & 40.880 & 0.9860  & 0.9858 & 64.90 \\
AHDRNet \cite{gong2019attenhdr} & 43.631 & 41.143 & 0.9900 & 0.9702 & 64.61 \\
NHDRRNet \cite{yan2020deep}\(^\dagger\) & 42.414 & - & 0.9887 & - & 61.21\\
Ours &  {\textbf{43.922}} &  {\textbf{41.572}} &  {\textbf{0.9905}} &  {\textbf{0.9865}} & {\textbf{65.45}} \\
    \bottomrule
	\end{tabular}
	\label{tab:compare_with_soa}
\end{table}

To evaluate our model, we compare it with the state-of-the-art HDR methods on the test images in the HDR dataset~\cite{LearningHDR_Kalantari}. The state-of-the-art methods used for comparison include two patch-based methods, Sen \textit{et al.}~\cite{sen2012} and Hu \textit{et al.} \cite{hu2013hdr}, and four deep-learning-based methods, Kalantari \textit{et al.} \cite{LearningHDR_Kalantari}, DeepHDR \cite{Wu_2018_ECCV_dhdr}, AHDRNet \cite{gong2019attenhdr}, and NHDRRNet \cite{yan2020deep}. Note that Kalantari \textit{et al.} \cite{LearningHDR_Kalantari} apply optical flow to align the input images in the preprocessing stage. DeepHDR \cite{Wu_2018_ECCV_dhdr} requires the background of the input images to be aligned by homography transformation, while AHDRNet \cite{gong2019attenhdr} and our method do not need to align the input images in the preprocessing stage.

Table \ref{tab:compare_with_soa} shows the results of the experiment. Note that the quantitative results for Sen \textit{et al.}~\cite{sen2012}, Hu \textit{et al.} \cite{hu2013hdr}, and DeepHDR \cite{Wu_2018_ECCV_dhdr} are from \cite{Wu_2018_ECCV_dhdr}; those for NHDRRNet \cite{yan2020deep} are from \cite{yan2020deep}.
On the other hand, we re-train the models from \cite{LearningHDR_Kalantari} and  \cite{gong2019attenhdr} and report the collected results. This is because we find that their reported results are slightly lower than the results that we obtain in running their models.
Nevertheless, we can see that our method outperforms all the comparison methods on all metrics. \wx{Our top-performing results in the terms of PSNR$_{_{L}}$ and PSNR$_{\mu}$ implies the generation capability of our proposed method in the HDR domain and the tonemapping domain. Compared with the latest works \cite{gong2019attenhdr,yan2020deep} that introduce attention schemes to relieve the LDR misalignment problem, their results in the terms of SSIM$_{\mu}$ and SSIM$_{L}$ are not on par with those of our approach, indicating that their models cannot consistently preserve the structural information of produced HDR images as ours. In addition, our results also demonstrate a healthy margin over other baselines on HDR-VDP-2 for the produced HDR image quality.}

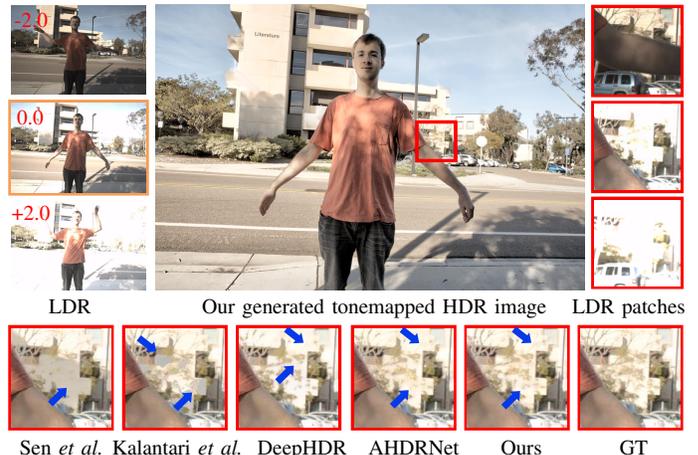
\begin{figure}
\footnotesize
    \centering
    \includegraphics[width=0.5\textwidth]{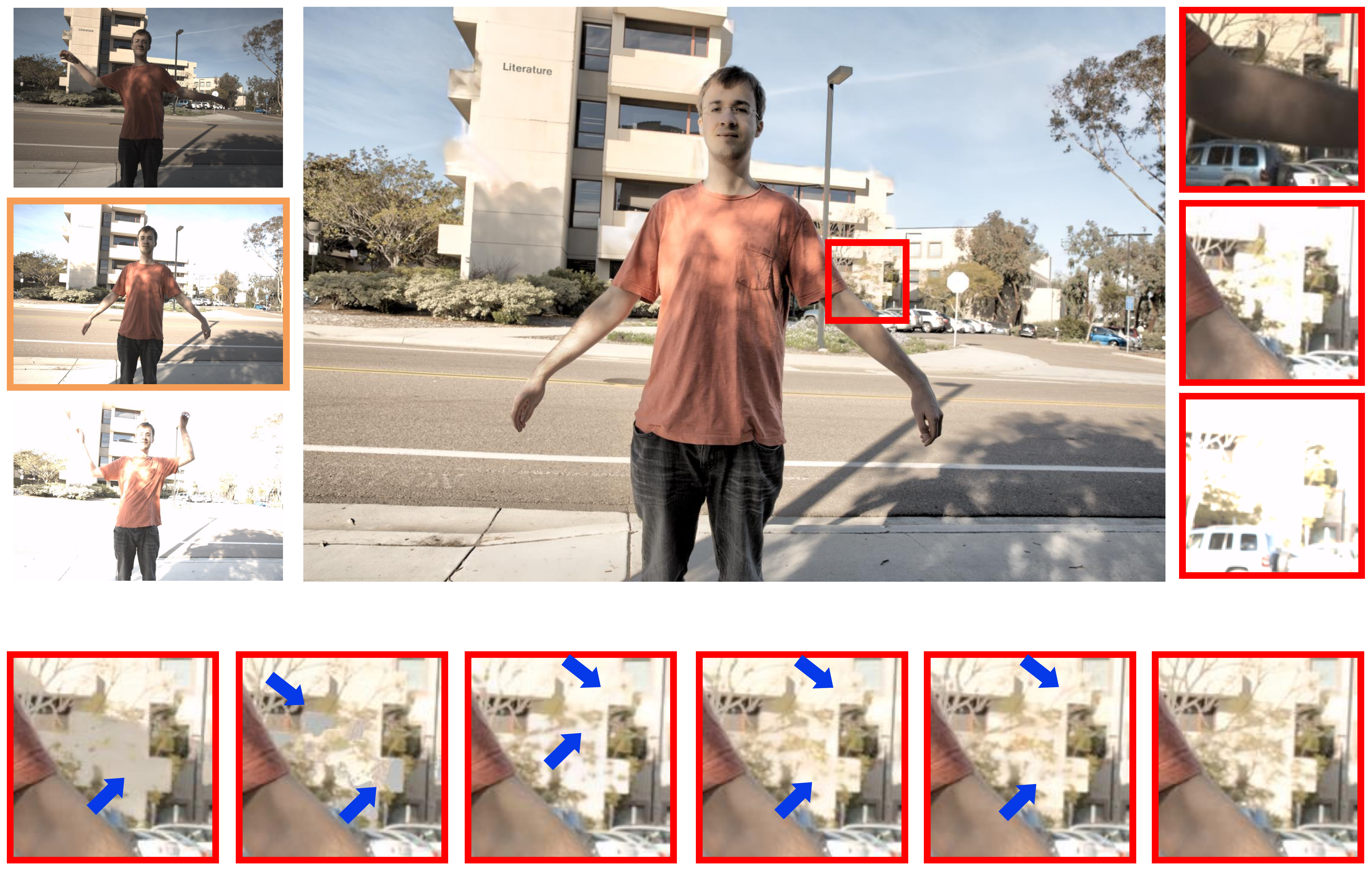}
    \begin{picture}(0,0)
    \put(-121, 163){\textcolor{red}{-2.0}}
    \put(-120, 126){\textcolor{red}{0.0}}
    \put(-122, 90){\textcolor{red}{+2.0}}
    \put(-108,54.5){LDR}
    \put(-50,54.5){Our generated tonemapped HDR image}
    \put(90,54.5){LDR patches}


    \put(-119,1){Sen \textit{et al.}}
    \put(-84,1){Kalantari \textit{et al.}}
    \put(-29,1){DeepHDR} 
    \put(13,1){AHDRNet} 
    \put(63,1){Ours}
    \put(108,1){GT}
    \end{picture}
    \vspace{-0.3cm}
    \caption{An example from the dataset~\cite{LearningHDR_Kalantari}. We compare a set of patches cropped from the tonemapped HDR images generated by the state-of-the-art methods. The blue arrows highlight the differences among the results of the comparison methods.}
    \label{fig:soa_5}
\end{figure}

\begin{figure}
\footnotesize
    \centering
    \includegraphics[width=0.5\textwidth]{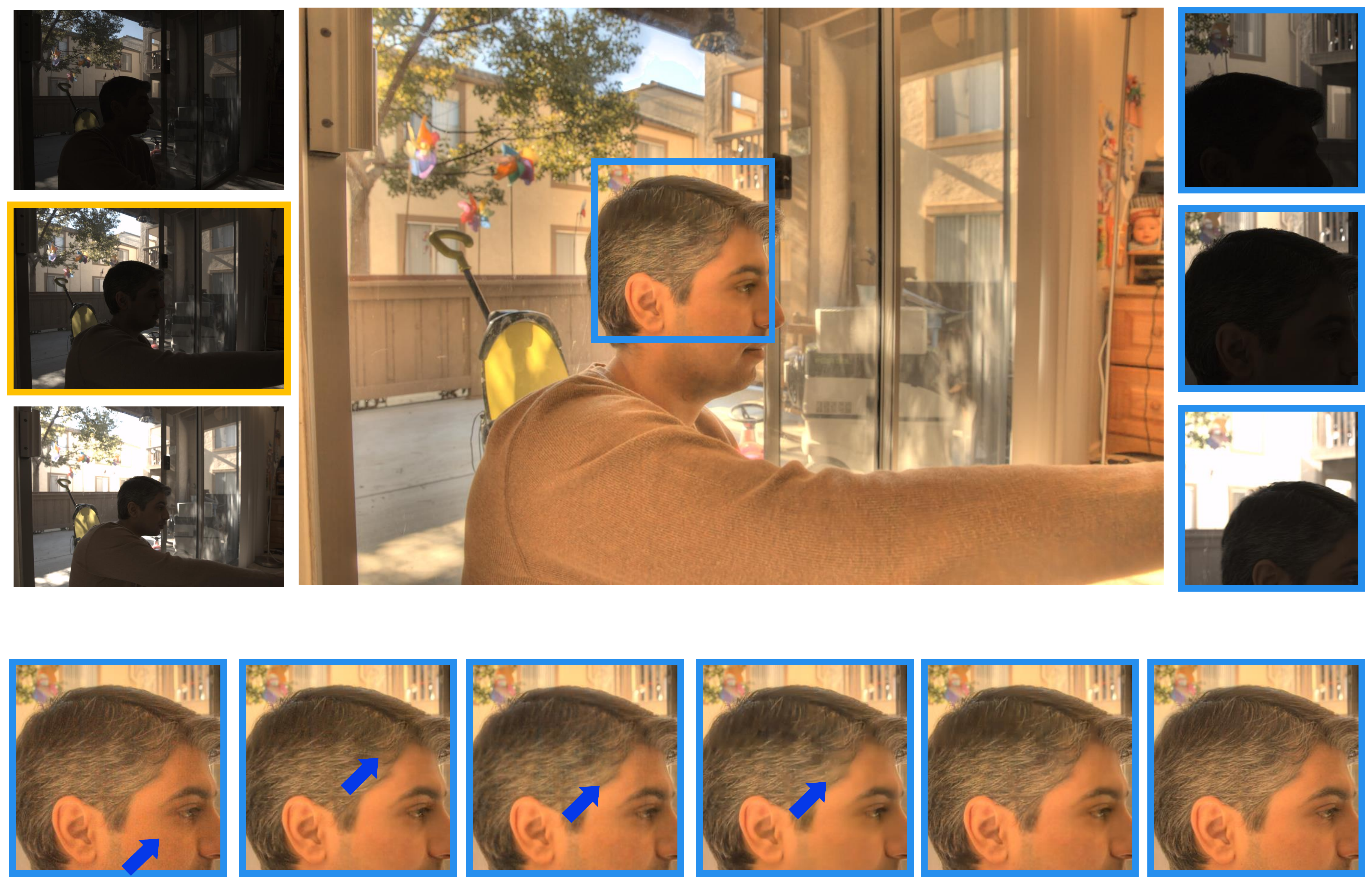}
    \begin{picture}(0,0)
    \put(-120, 164){\textcolor{red}{-2.0}}
    \put(-120, 128){\textcolor{red}{0.0}}
    \put(-120, 91){\textcolor{red}{+2.0}}
    \put(-108,55.5){LDR}
    \put(-50,55.5){Our generated tonemapped HDR image}
    \put(90,55){LDR Patches}

    \put(-118,1.5){Sen \textit{et al.}}
    \put(-83,1){Kalantari \textit{et al.}}
    \put(-26,1){DeepHDR} 
    \put(16,1){AHDRNet} 
    \put(65,1){Ours}
    \put(108,1){GT}
    \end{picture}
    \vspace{-0.2cm}
    \caption{{Another example from the dataset~\cite{LearningHDR_Kalantari}.} We compare a set of patches cropped from the tonemapped HDR images generated by the state-of-the-art methods. The blue arrows highlight the differences among the results of the comparison methods.}
    \label{fig:soa_1}
\end{figure}

\begin{figure}
    \vspace{-0.5cm}
\footnotesize
    \centering
    \includegraphics[trim=0 0 0 0,clip, width=0.5\textwidth]{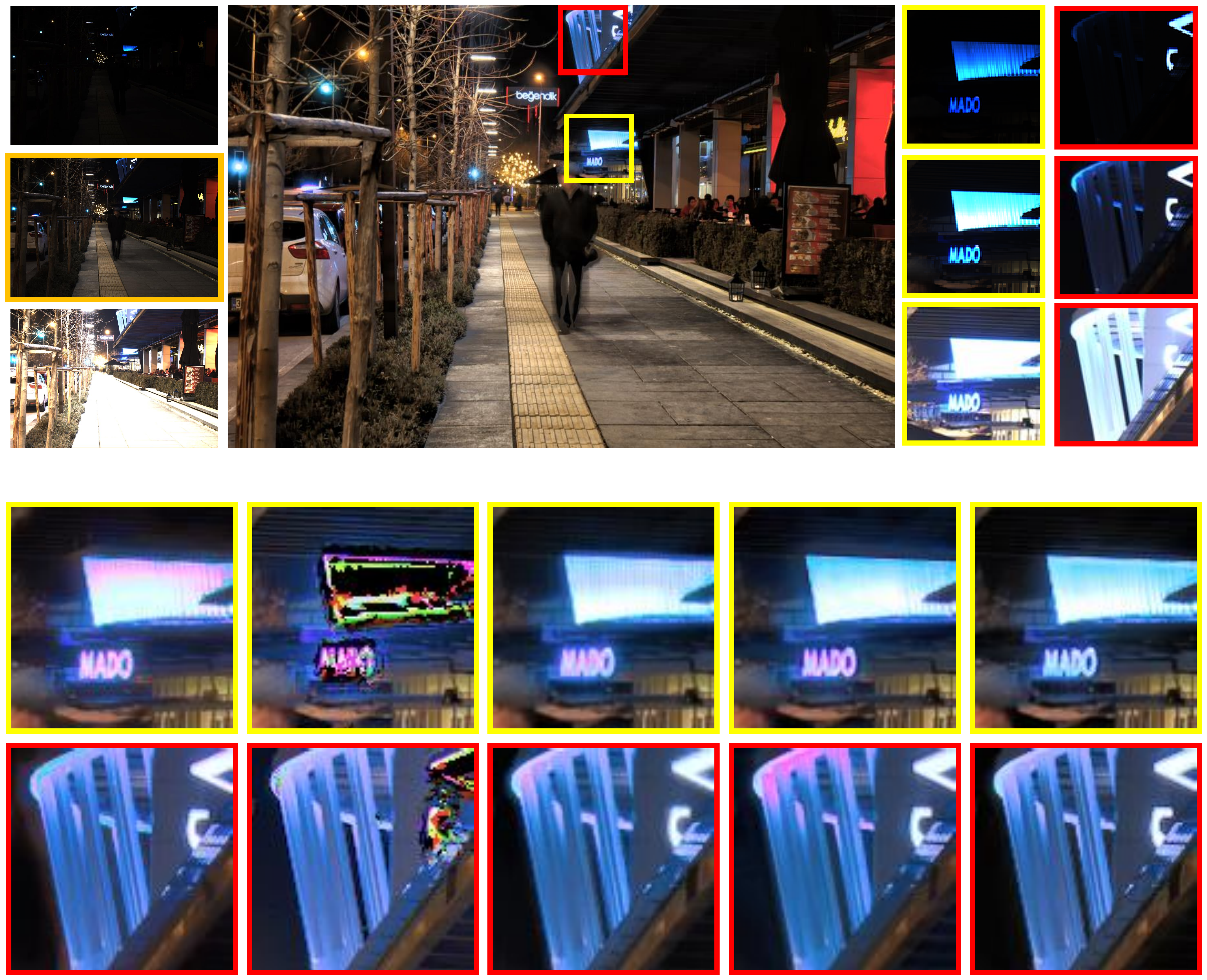}

    \begin{picture}(0,0)

    \put(-120, 208.5){\textcolor{red}{-3.0}}
    \put(-120, 177){\textcolor{red}{0.0}}
    \put(-120, 145){\textcolor{red}{+3.0}}
    \put(-108,114){LDRs}
    \put(-68,114){Our generated tonemapped HDR image}
    \put(80,114){LDR Patches}

    \put(-112,1){Sen \textit{et al.}}
    \put(-73,1){Kalantari \textit{et al.}}
    \put(-9,1){DeepHDR} 
    \put(43,1){AHDRNet} 
    \put(102,1){Ours}
    \end{picture}
    \vspace{-0.2cm}
    \caption{An example from another dataset~\cite{tursun2016} (which does not provide ground-truth images). We qualitatively compare two sets of patches cropped from the tonemapped HDR images generated by the state-of-the-art methods.}
    \label{fig:soa_3}
\end{figure}
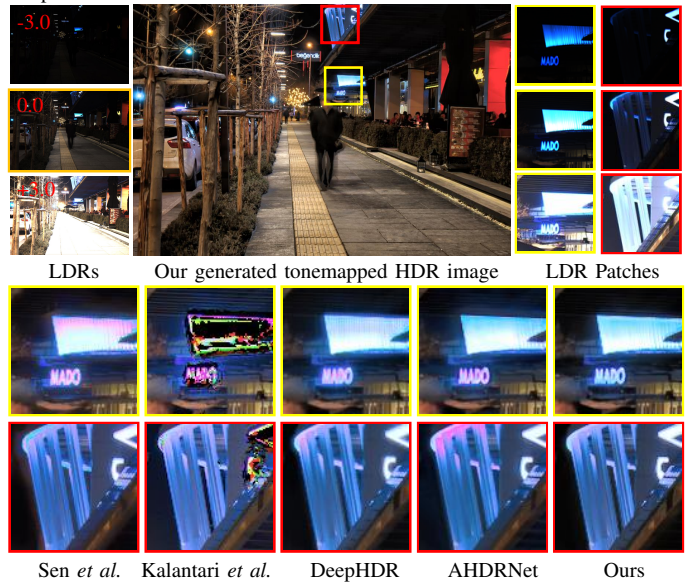

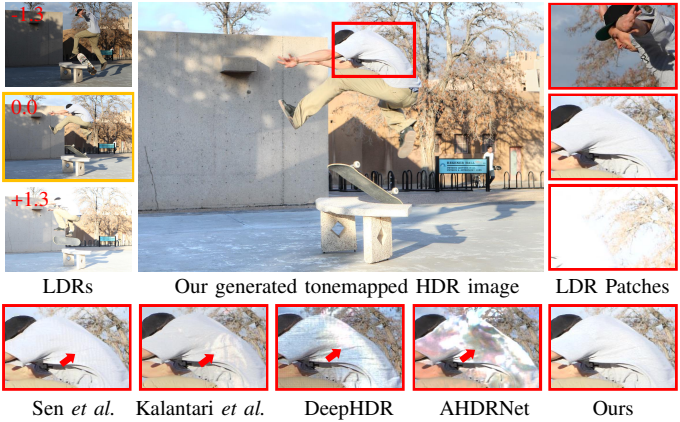
\begin{figure}
\footnotesize
    \centering
    \includegraphics[width=0.5\textwidth]{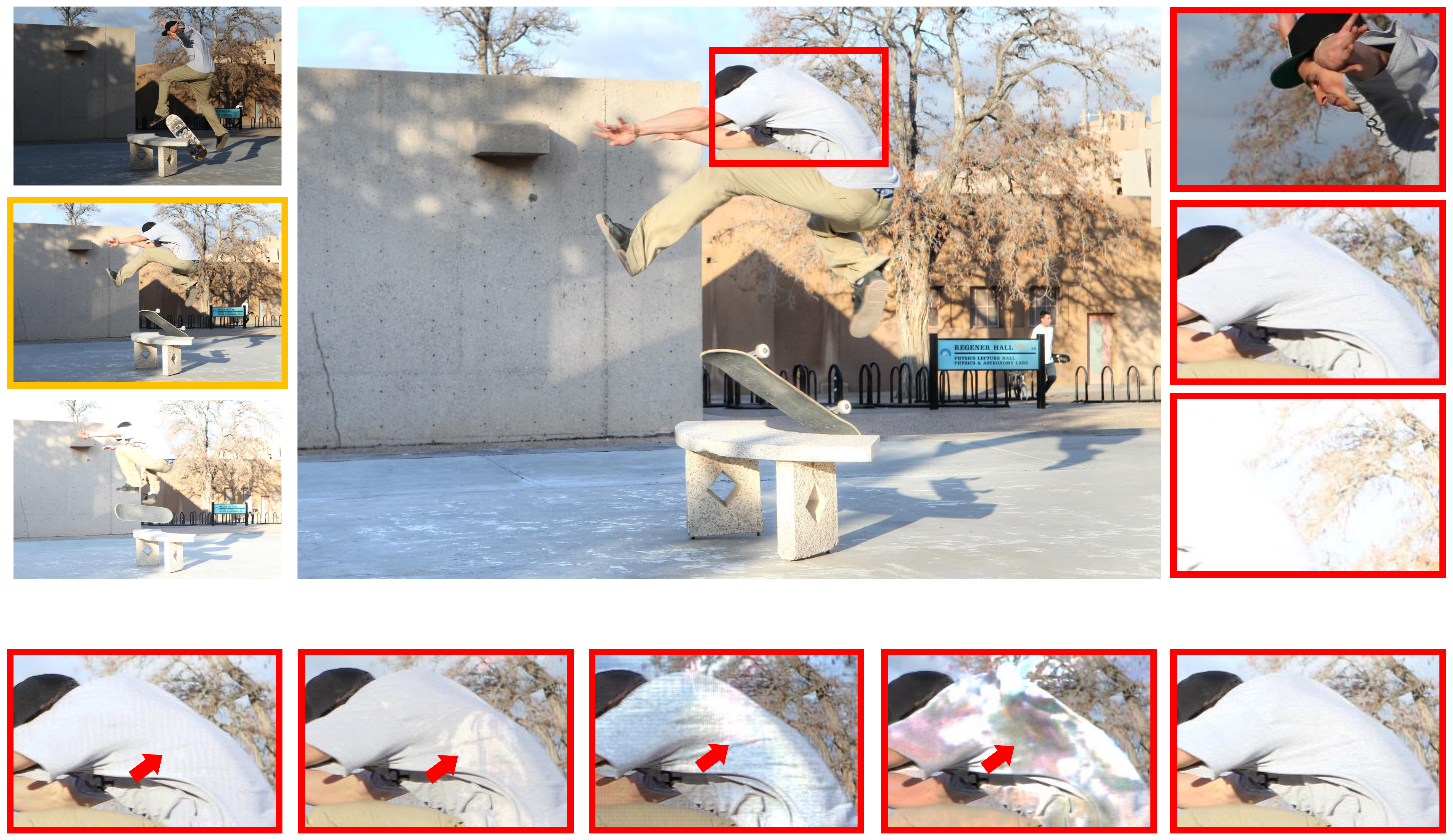}
    \begin{picture}(0,0)

    \put(-120, 150){\textcolor{red}{-1.3}}
    \put(-120, 115){\textcolor{red}{0.0}}
    \put(-120, 80){\textcolor{red}{+1.3}}
    \put(-108,47){LDRs}
    \put(-58,47){Our generated tonemapped HDR image}
    \put(86,47){LDR Patches}

    \put(-112,1){Sen \textit{et al.}}
    \put(-73,1){Kalantari \textit{et al.}}
    \put(-9,1){DeepHDR} 
    \put(42,1){AHDRNet} 
    \put(100,1){Ours}
    \end{picture}
    \vspace{-0.3cm}
    \caption{An example from yet another dataset~\cite{sen2012} (which does not provide ground-truth images). We qualitatively compare one set of patches cropped from the tonemapped HDR images generated by the state-of-the-art methods. The red arrows highlight the differences among the results of the comparison methods.}
    \label{fig:soa_4}
\end{figure}

We also show several example scenarios with dynamic objects or camera motions from public datasets for visual comparison, as shown in Figs. \ref{fig:soa_5}, \ref{fig:soa_1}, \ref{fig:soa_3} and \ref{fig:soa_4}. Note that the HDR images are tonemapped using \textit{Photomatix}\footnote{https://www.hdrsoft.com}, which is a tonemapper different from the one used in training. Figs.~\ref{fig:soa_5} and \ref{fig:soa_1} show that \wx{our model is capable of recovering the image details, such as hair and sticks, from the LDR images, thanks to the LDR alignment in the feature domain.} Specifically, Sen \textit{et al.}~\cite{sen2012} tends to introduce noise to the generated images (e.g., the face in Fig.~\ref{fig:soa_1}), and the other methods may produce various degrees of blurry artifacts. \wx{On the contrary, our GAN-based method is able to generate more faithful information.}
Fig.~\ref{fig:soa_3} shows an image with a walking person and global camera motion, under a low-light condition. While the other methods may produce HDR images with distorted colors, which is not consistent with the reference image, due to the misalignment, \wx{our method still can well align LDR images and performs well in such a challenging scene.} Fig.~\ref{fig:soa_4} shows a fast moving person, artifacts can be easily observed in the generated HDR images from all existing methods. In contrast, our result is much better, without the artifacts, indicating the effectiveness of our LDR feature alignment and the GAN-based paradigm of our model.

\subsection{Ablation Study}
\label{sec:ablation}

In this subsection, we analyze the effectiveness of different parts of our model.


\begin{table}
	\centering
	\footnotesize
	\setlength{\tabcolsep}{2pt}
	\caption{Ablation study of different variants of the proposed network structure on the four metrics.
	\textit{D.S.} indicates whether the deep supervision is adopted in a multi-scale network.
	The values highlighted in red and blue indicate the best and the second-best performances, respectively.}
	\vspace{-0.3cm}
	\begin{tabular}{c|c|c||cc|cc}
		\toprule
		Structure  & \textit{D.S.} & Output & PSNR$_\mu$ & PSNR$_L$ & SSIM$_\mu$ & SSIM$_L$ \\
		\midrule
		Two-scale & - &$\hat{H}_1$ & 43.710 & 41.139 & 0.9904 & 0.9855 \\
		\midrule
		&  &$\hat{H}_1$ & 43.847 & 41.318 & \textcolor{red}{\textbf{0.9906}} & 0.9860 \\
		Four-scale & $\checkmark$ &$\hat{H}_2$& \textcolor{blue}{\textbf{43.901}} & \textcolor{blue}{\textbf{41.416}} & \textcolor{red}{\textbf{0.9906}} & \textcolor{blue}{\textbf{0.9865}} \\
		&  &$\hat{H}_3$ & 43.833 & 41.363 & \textcolor{red}{\textbf{0.9906}} & \textcolor{red}{\textbf{0.9867}} \\
		\midrule
		w/o Merging blocks & - &$\hat{H}_1$ & 43.595 & 41.009 & 0.9900 & 0.9854 \\
		\midrule
		\multirow{3}{*}{Ours} & \multirow{2}{*}{$\checkmark$} &$\hat{H}_1$ & 43.787 & 41.223 & 0.9904 & 0.9853 \\
		&  &$\hat{H}_2$ & \textcolor{red}{\textbf{43.922}} & \textcolor{red}{\textbf{41.572}} & \textcolor{blue}{\textbf{0.9905}} & \textcolor{blue}{\textbf{0.9865}} \\
		\cmidrule{2-7}
		& $\times$ &$\hat{H}_2$ & 43.838 & 41.381 & \textcolor{blue}{\textbf{0.9905}} & 0.9857 \\
		\bottomrule
	\end{tabular}
	\label{tab:ablation2}
\end{table}

\begin{figure}
	\centering
	\footnotesize
	\begin{tabular}{cc}
		\includegraphics[width=0.18\textwidth]{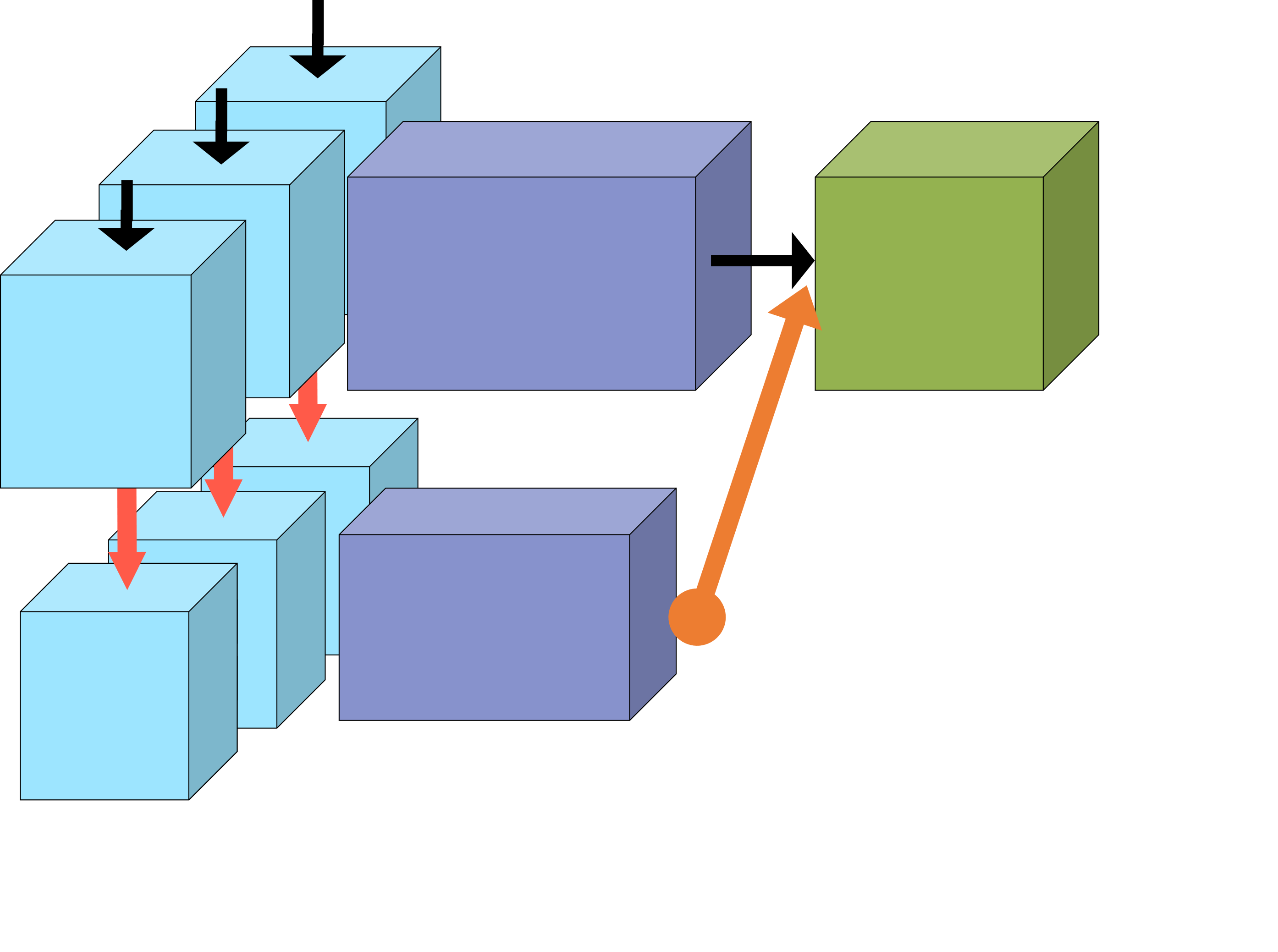} &
		\includegraphics[width=0.23\textwidth]{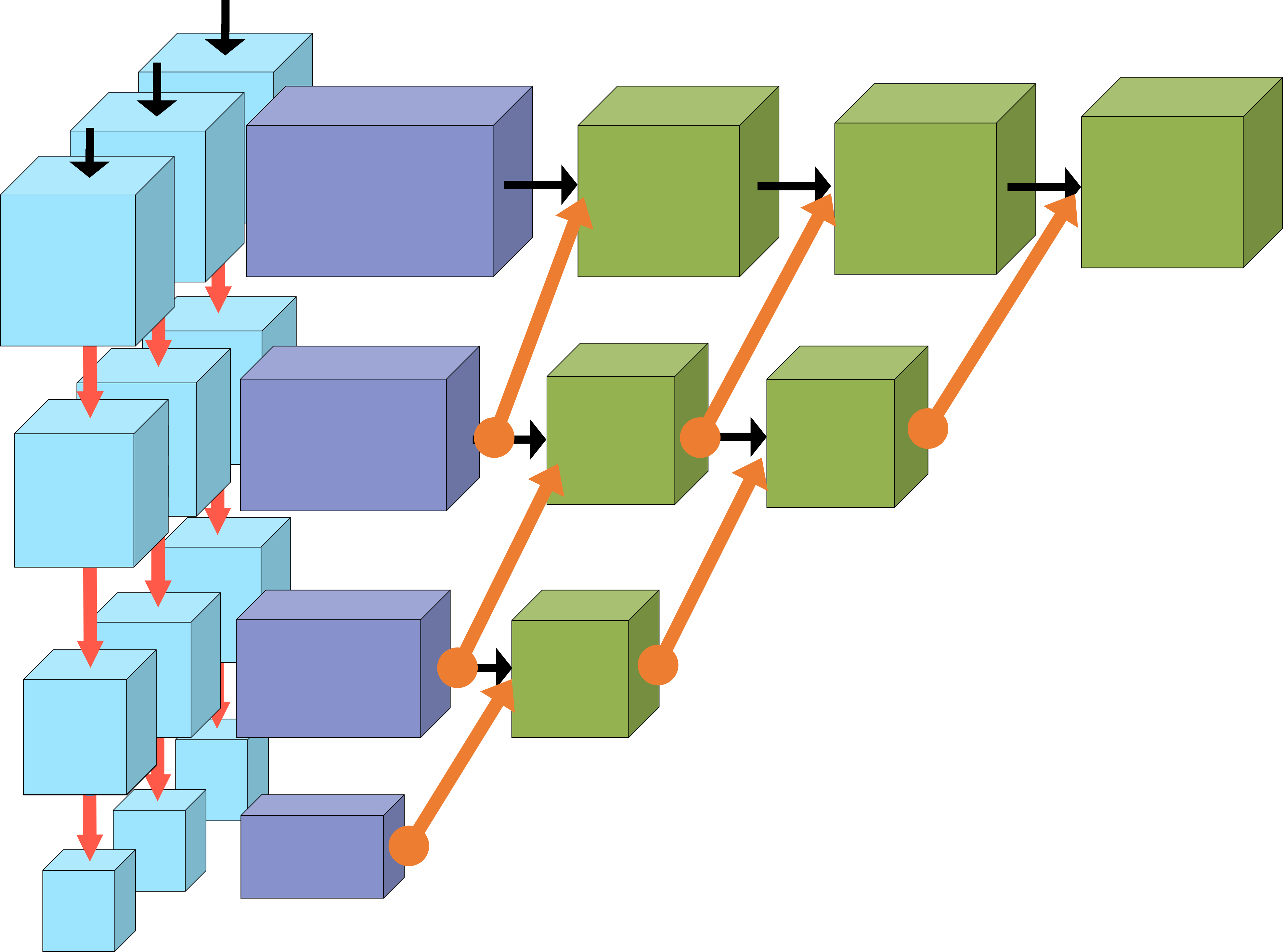}\\
		(a) Two-scale & (b) Four-scale\\
		\includegraphics[width=0.21\textwidth]{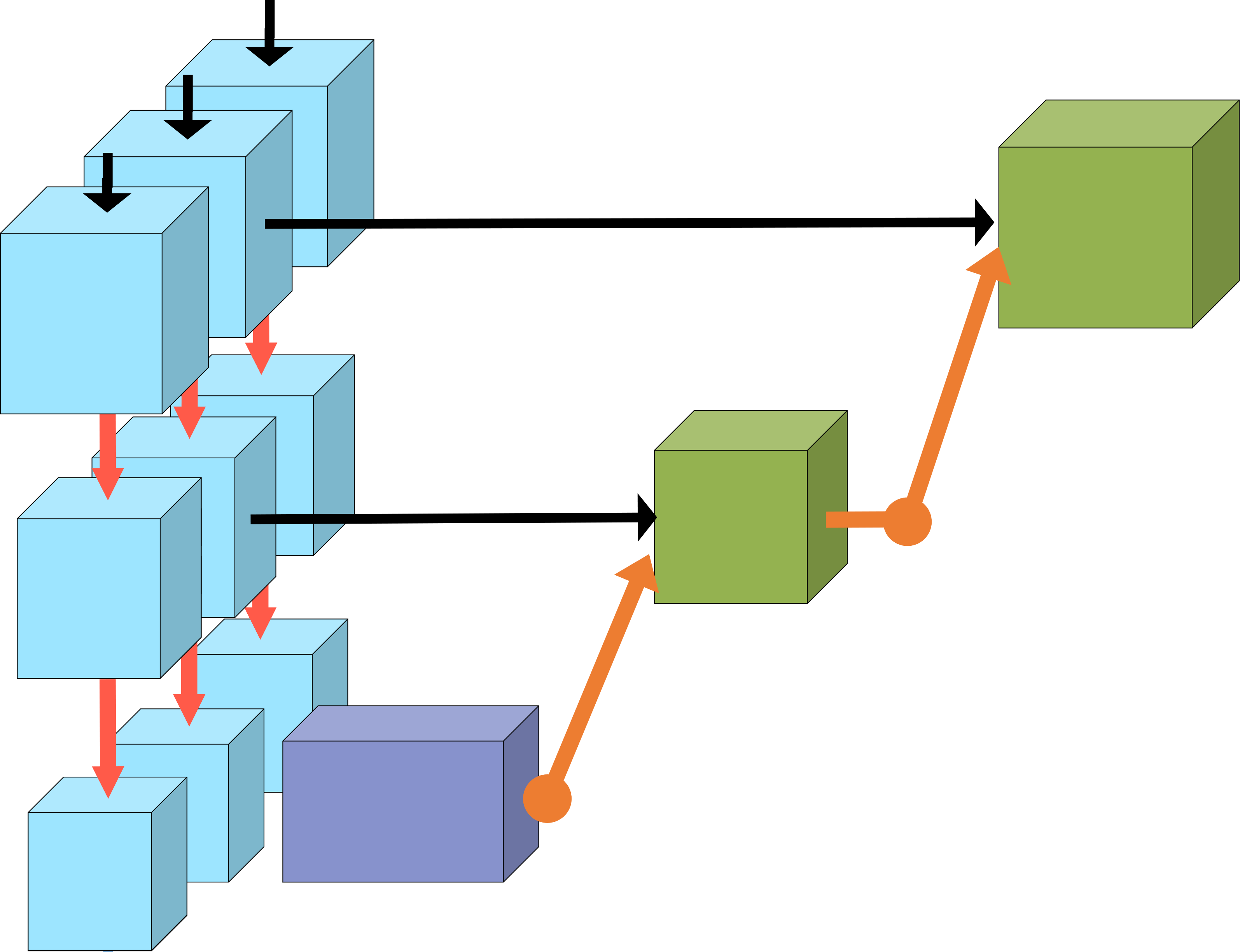} &
		\includegraphics[width=0.2\textwidth]{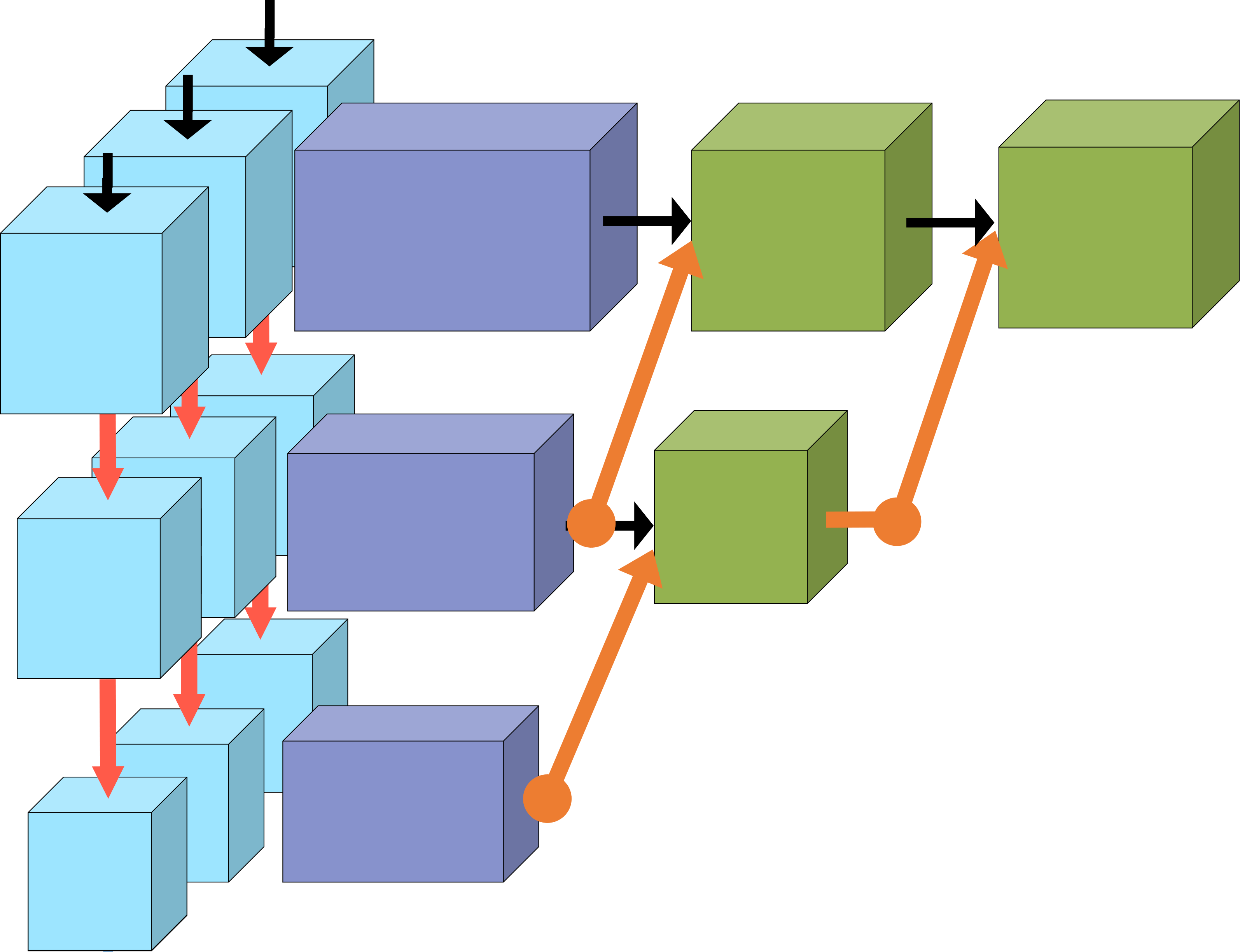}\\
		(c) w/o Merging blocks & (d) Ours
	\end{tabular}
	\vspace{-0.3cm}
	\caption{Different variants of our network structure, including (a) two-scale and (b) four-scale, (c) the network structure without the \wx{merging blocks}, and (d) our proposed model.}
	\label{fig:network_structure}
\end{figure}

\wx{\textbf{Network scale.}} To dissect the structure of our proposed generator, \wx{we first analyze the scale of its network structure. We introduce several variants for comparison, as shown in Fig.~\ref{fig:network_structure}, including a two-scale version (Fig.~\ref{fig:network_structure}(a)), a four-scale version (Fig.~\ref{fig:network_structure}(b)), and our proposed three-scale network (Fig.~\ref{fig:network_structure}(d)).}
Note that the two-scale network contains only one output branch for supervision (i.e., $\hat{H}_1$), while the four-scale network contains three output branches (i.e., $\hat{H}_1$, $\hat{H}_2$ and $\hat{H}_3$). Table~\ref{tab:ablation2} shows the corresponding results (i.e., first, second and fourth groups of results).
Intuitively, a shallower network may not be able to capture sufficient multi-scale features for the task, while a deeper network can cause redundancy, network learning difficulties or even overfitting.
We can observe from Table~\ref{tab:ablation2} that the two-scale network does not perform well, while the deeper network has slightly lower performances, compared with our proposed three-scale network. It is worth noting that the best performing output branch of the four-scale network is the second one, which implies that the last output branch of the network may be overfitted during training. From these results, we may conclude that the three-scale architecture is the optimal one.

\begin{figure}
	\footnotesize
	\centering
	\includegraphics[trim=0 0 0 0,clip, width=0.5\textwidth]{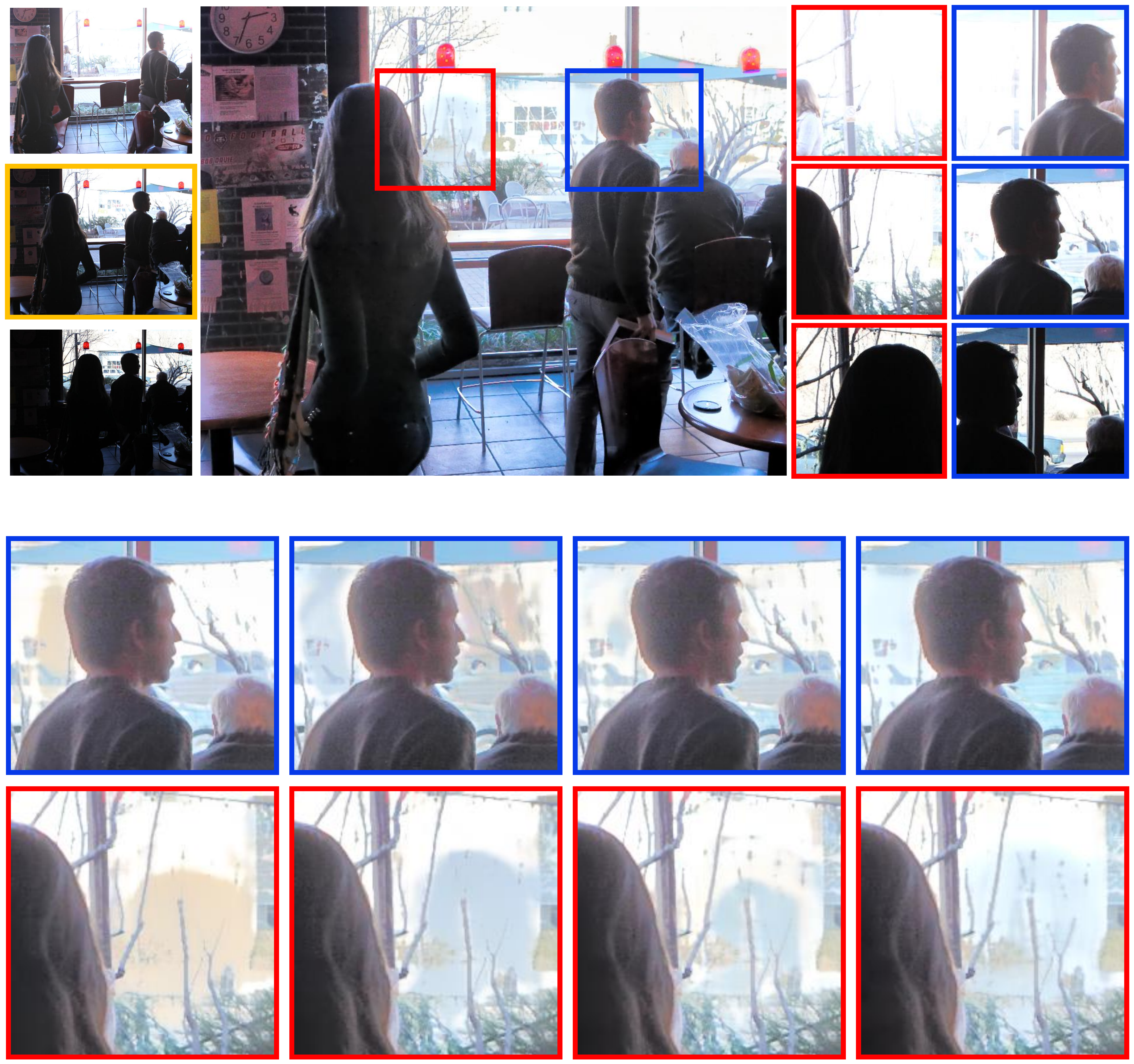}
	\begin{picture}(0,0)
	\put(-135,0){(a) w/o Merging blocks}
	\put(-51,0){(b) Two-scale}
	\put(13,0){(c) Four-scale}
	\put(86,0){(d) Ours}
	
	\put(-121, 242){\textcolor{red}{+2.0}}
	\put(-121, 205){\textcolor{red}{0.0}}
	\put(-121, 168){\textcolor{red}{-2.0}}
	\put(-110,133.5){LDRs}
	\put(-78, 133.5){Our generated tonemapped HDR image}
	\put(67,133.5){LDR patches}
	\end{picture}
	
	\vspace{-0.2cm}
	\caption{An example to compare the visual results of different variants of our network structure.}
	\label{fig:ablation_struction}
\end{figure}
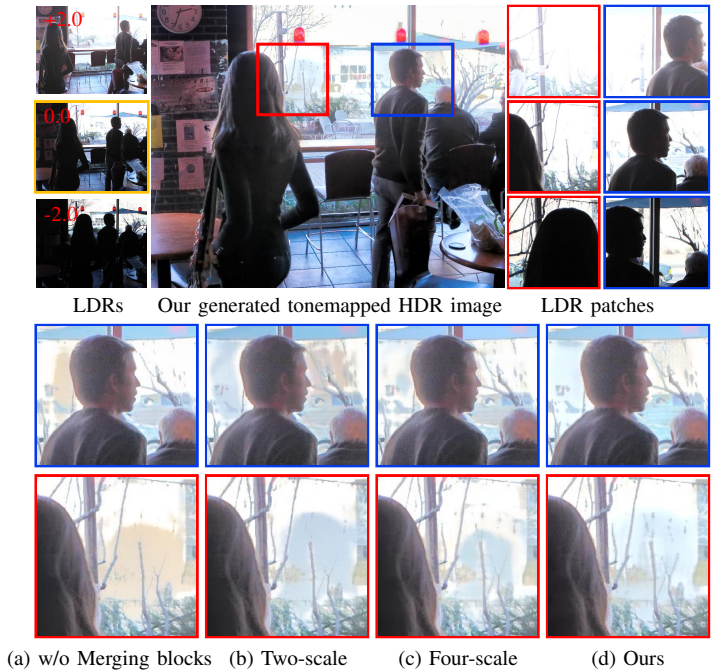

Fig.~\ref{fig:ablation_struction} shows two patches of an example image, and the results of these two patches from the three variants. In this challenging scene, the background is over-exposed and partially occluded due to the motions of the two persons. We can see that both the two-scale network (Fig.~\ref{fig:ablation_struction}(b)) and the four-scale network (Fig.~\ref{fig:ablation_struction}(c)) have obvious artifacts. In contrast, our model (Fig.~\ref{fig:ablation_struction}(d)) produces much better results.

\wx{\textbf{Merging blocks.}}
We then study the effectiveness of the merging blocks. As mentioned, to align the features from multiple exposures and preserve the details of the generated images, we densely align the features via the residual merging blocks.
To validate this, we compare our network with the network without the merging blocks as shown in Fig.~\ref{fig:network_structure}(c). The comparison results are also shown in Table~\ref{tab:ablation2}. The model w/o merging blocks (third groups of results in Table~\ref{tab:ablation2}) has a performance significantly lower than the proposed model (fourth groups of results in Table~\ref{tab:ablation2}), in terms of PSNRs. We can also observe from the visual results in Fig.~\ref{fig:ablation_struction}(a) that without the merging blocks, the model fails to align the three input images and is thus unable to recover the contents well.

\begin{figure}
    \centering
    \footnotesize
    \includegraphics[width=0.5\textwidth]{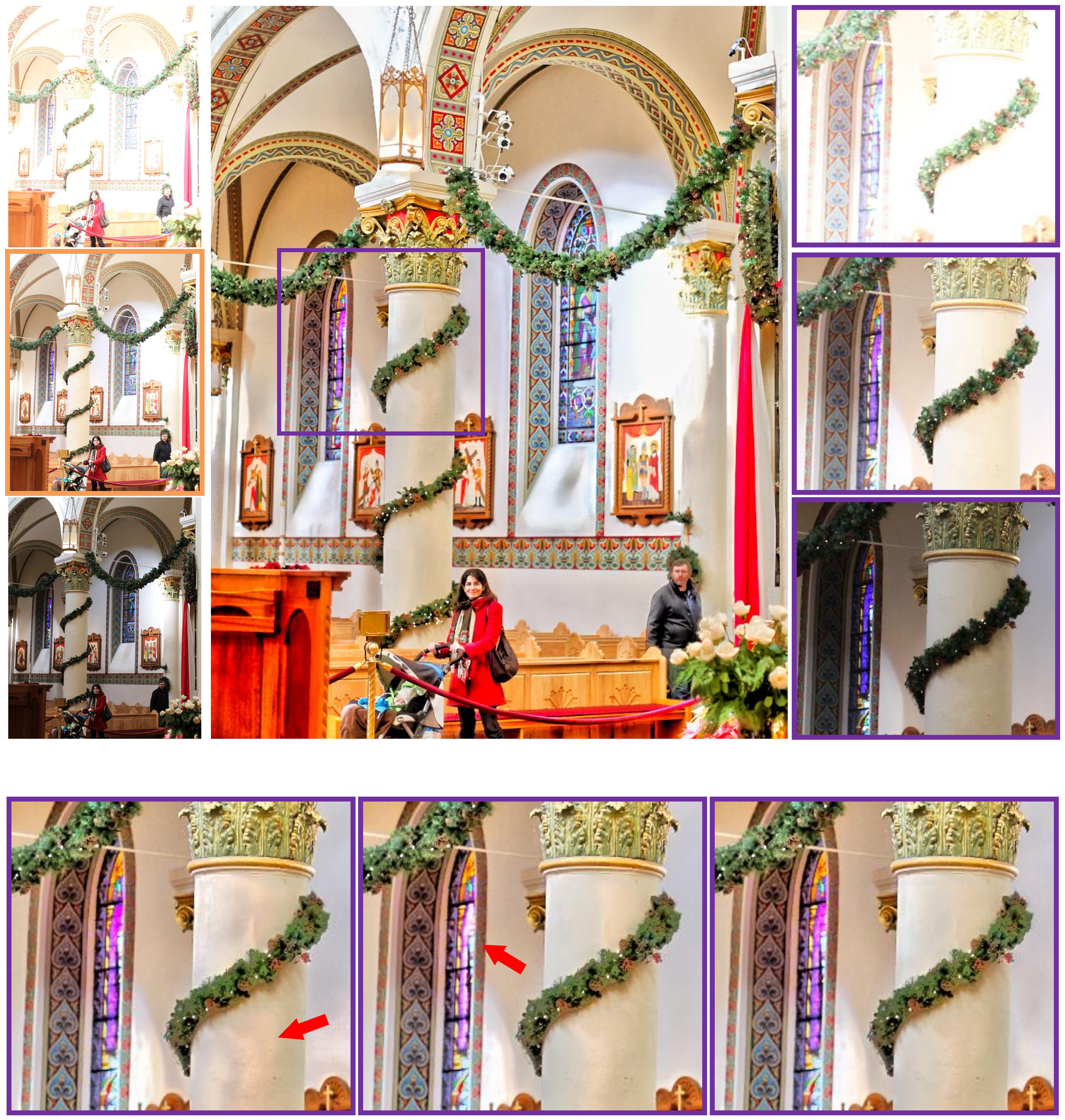}
    \begin{picture}(0,0)
    \put(-121,272){\textcolor{red}{+4.0}}
    \put(-121, 212){\textcolor{red}{+2.0}}
    \put(-121, 152){\textcolor{red}{0.0}}
    \put(-109,92){LDR}
    \put(-70,92){Our generated tonemapped HDR image}
    \put(76,92){LDR patches}

    \put(-115,0){(a) Only $\hat{H}_2$ output}
    \put(-17,0){(b) $\hat{H}_1$ output}
    \put(62,0){(c) $\hat{H}_2$ output}
    \end{picture}
    \vspace{-0.2cm}
    \caption{We show example HDR images generated by using (a) using $\hat{H}_2$ as the model output and without using $\hat{H}_1$ in model training, (b) using $\hat{H}_1$ as the model output, and (c) using $\hat{H}_2$ as the model output.}
    \label{fig:ablation_3}
\end{figure}
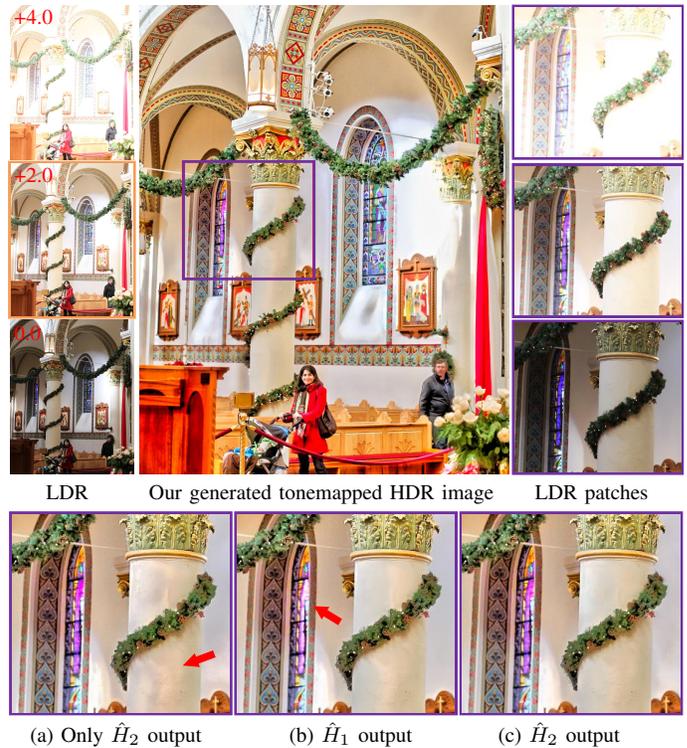

\textbf{Deep HDR supervision.}
To demonstrate the effectiveness of our deep HDR supervision, we further conduct several experiments for a thorough analysis.
First, we evaluate the generated HDR results with and without deep supervision on the metrics. In addition, we also compare these results with the results from the network trained using only the second output branch (i.e., $\hat{H}_1$ is discarded and only $\hat{H}_2$ is used in the loss function). The last groups of results in  Table~\ref{tab:ablation2} shows these results. As observed, in our proposed model, \wx{the result of $\hat{H}_2$ outperforms that of $\hat{H}_1$, since the information from $\hat{H}_1$ helps further refine $\hat{H}_2$} by compensating with more details to the generated $\hat{H}_2$. For reference, given the model training with \wx{$\hat{H}_2$} only, it performs worse than the one with multiple supervisions. We show a visual example in Fig.~\ref{fig:ablation_3}, which demonstrates the results from different output branches. We can see the best result is produced when using both $\hat{H}_1$ and $\hat{H}_2$ in the loss function and with $\hat{H}_2$ as the output, which is consistent with our quantitative results.

\begin{figure}
	\centering
	\footnotesize
	\includegraphics[width=0.4\textwidth]{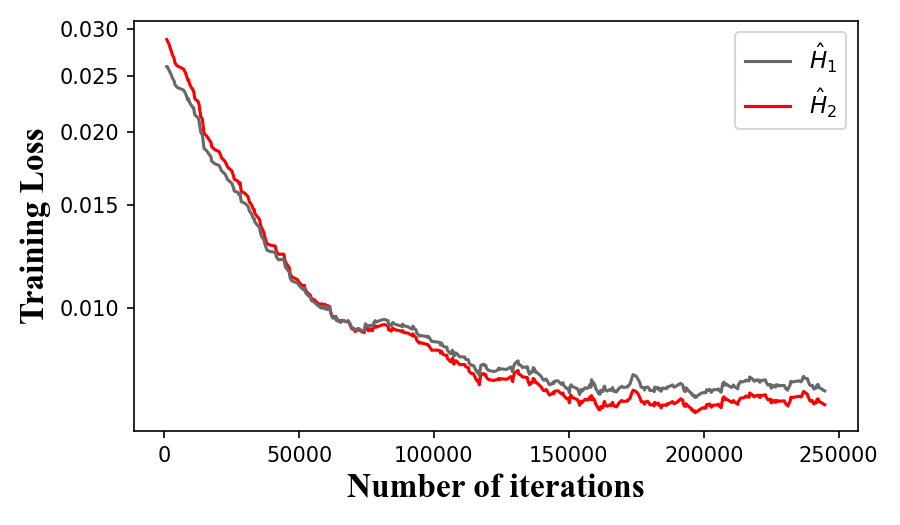}
	\vspace{-0.2cm}
	\caption{Training losses of the two output branches ($\hat{H}_1$ and $\hat{H}_2$).}
	\label{fig:two_output_train}
\end{figure}

Second, we show the training losses of our proposed model in Fig. \ref{fig:two_output_train}. The gray curve indicates the training loss of $\hat{H}_1$, while the red curve indicates the training loss of $\hat{H}_2$. 
We can see from the results that when the number of iterations is small ($<$ 40,000), $\hat{H}_1$ performs better. This is because $\hat{H}_1$ has a shallower structure. However, with a higher number of iterations ($>$ 70,000), $\hat{H}_2$ performs comparatively better.

{}


\begin{table}
	\centering
	\footnotesize
	\setlength{\tabcolsep}{2pt}
	\caption{Ablation study on the effectiveness of our discriminator $D$. We first compare against the framework without the discriminator (i.e., `w/o $D$'). We also compare with the discriminator which takes only one generated HDR image ($\hat{H}_1$ or $\hat{H}_2$) and the ground-truth image as inputs. The values highlighted in red and blue indicate the best and the second-best performance, respectively.}
	\vspace{-0.3cm}
	\begin{tabular}{c|c||cc|cc}
\toprule
Structure &  Output & PSNR$_\mu$ & PSNR$_L$ & SSIM$_\mu$ & SSIM$_L$ \\
\midrule
\multirow{2}{*}{w/o $D$} & $\hat{H}_1$&43.699 & 41.005 & 0.9903 & 0.9857 \\
 & $\hat{H}_2$ &43.729 & 41.013 & 0.9903 & 0.9858 \\
 \midrule
w/ $D$ & $\hat{H}_1$ & 43.634 & 41.132 & 0.9902 & 0.9854 \\
(w/o $\hat{H}_2$ in $\mathcal{L}_{adv}$) & $\hat{H}_2$ & 43.693 & 41.085 & 0.9903 & 0.9857 \\
\midrule
w/ $D$ & $\hat{H}_1$ & 43.535 & 40.991 & 0.9901 & 0.9851 \\
(w/o $\hat{H}_1$ in $\mathcal{L}_{adv}$) & $\hat{H}_2$ & 43.756 & \textcolor{blue}{\textbf{41.281}} & 0.9903 & \textcolor{blue}{\textbf{0.9861}} \\
\midrule
\multirow{2}{*}{Ours} & $\hat{H}_1$ & \textcolor{blue}{\textbf{43.787}} & 41.223 & \textcolor{blue}{\textbf{0.9904}} & 0.9853 \\
& $\hat{H}_2$ & \textcolor{red}{\textbf{43.922}} & \textcolor{red}{\textbf{41.572}} & \textcolor{red}{\textbf{0.9905}} & \textcolor{red}{\textbf{0.9865}} \\
		\bottomrule
	\end{tabular}
	\label{tab:ablation1}
\end{table}

\wx{\textbf{Discriminator.}} To evaluate the effectiveness of our discriminator $D$ for generating HDR images, we first compare the results from our model without the discriminator $D$. In particular, as observed in Table~\ref{tab:ablation1}, comparing the models with and without $D$, the performance of our model with $D$ is apparently boosted. It is interesting to mention that our generator network itself (i.e., without introducing the discriminator nor the GAN loss) can achieve comparable performance with the state-of-the-art methods. Specifically, we can observe from Tables~\ref{tab:compare_with_soa} and \ref{tab:ablation1} that our trained generator without GAN obtains a better result in terms of PSNR$_{\mu}$ (i.e., $43.729$) than AHDRNet (i.e., $43.631$), while our model with GAN obtains an even better result (i.e., $43.922$).



\wx{In addition, we also analyze the effectiveness of using multiple generated HDR images in the discriminator. Recall that during training, our discriminator $D$ receives two deep supervised results from the generator (i.e., $\hat{H}_1$ and $\hat{H}_2$) as well as the ground-truth image (i.e., $H_{gt}$) as inputs.
%
These two deep supervised results are supposed to be similar, but not exactly the same. To evaluate if it is necesary to have both of them as inputs to the discriminator, we train our GAN-based model with $\hat{H}_1$ or $\hat{H}_2$ only. In other words, according to Eq.~\ref{eq:adv_loss}, only the first output $\hat{H}_1$ or the second output $\hat{H}_2$ from $G(\cdot)$ is used in the adversarial loss $\mathcal{L}_{adv}$.
Table~\ref{tab:ablation1} shows the results, denoted as ``w/ $D$ (w/o $\hat{H}_2$ in $\mathcal{L}_{adv}$)" and ``w/ $D$ (w/o $\hat{H}_1$ in $\mathcal{L}_{adv}$)". From these results, we can observe that sending two outputs to the discriminator helps improve the network generation ability slightly. This is because $D$ needs to discriminate not only $\hat{H}_2$ but also $\hat{H}_1$ from the ground-truth, which thus encourages the model to improve the quality of the generated HDR images. }


\section{Conclusion and Future Works}
\label{sec:conclusion}

In this paper, we propose a novel  GAN-based HDR model, \textit{HDR-GAN}, to address the problems caused by large object motions in the scene. Given three multi-exposed LDR images as input, our model generates an HDR image. By incorporating the adversarial learning scheme, our method is able to produce faithful information in the regions with missing contents. In addition, we also propose a novel generator network, with reference-based residual merging blocks for aligning large object motions in the feature domain, and a deep HDR supervision scheme for eliminating artifacts of the reconstructed HDR images. Our extensive experiments demonstrate that the proposed model can obtain state-of-the-art reconstruction performance, compared with the prior HDR methods.

\wx{Generally, HDR imaging is still a challenging research problem, especially when the reference image of the input LDR images suffer from significant quality degeneration, e.g., severe missing content due to under- or over-saturation. Thus, existing datasets may not be sufficient to push forward the progress of HDRI techniques, since the HDR dataset \cite{LearningHDR_Kalantari} only contains images from a limited number of less challenging scenes and its performance measures for the state-of-the-art methods are hard to be further improved, while the other datasets do not provide ground-truth. As a future work, it will be necessary to establish a larger HDR dataset with more diverse and challenging scenes. Besides, since the resolution of photos taken by the latest electronic products become very high, directly reconstructing HDR images from such high-resolution images is a formidable task for the off-the-shelf CNN-based models due to the limited memory resources. In the future, it is needed to investigate a model that can efficiently and effectively perform HDR imaging for high-resolution images.}

\bibliographystyle{IEEEtran}
\bibliography{references}

\ifCLASSOPTIONcaptionsoff
  \newpage
\fi

\end{document}